\documentclass[%
 reprint,
 amsmath,amssymb,
 aps,
]{revtex4-2}
\usepackage{CJK}
\usepackage{graphicx}
\usepackage{dcolumn}
\usepackage{bm}
\usepackage{amsmath}

\begin{document}
\preprint{APS/123-QED}
\title{Genetics-based deperturbation analysis for the spin-orbit coupled ${\mathrm A}^1\Sigma^+$ and ${\mathrm b}^3\Pi_{0^+}$ states of LiRb\\}

\author{Yide Yin}
	\affiliation{School of Physics, Dalian University of Technology, 116024 Dalian, China}
\author{Xuhui Bai}
\affiliation{School of Physics, Dalian University of Technology, 116024 Dalian, China}
\author{Xuechun Li}
\affiliation{School of Physics, Dalian University of Technology, 116024 Dalian, China}
\author{Xin-Yu Luo}
\affiliation{Max Planck Institute of Quantum Optics, 85748 Garching, Germany}
\affiliation{Munich Center for Quantum Science and Technology, 80799 München, Germany}

\author{Jie Yu}%
	\email{yujie@dlut.edu.cn}
	\affiliation{School of Physics, Dalian University of Technology, 116024 Dalian, China}
\author{Gaoren Wang}%
    \email{gaoren.wang@dlut.edu.cn}
	\affiliation{School of Physics, Dalian University of Technology, 116024 Dalian, China}
\author{Yongchang Han}
	\affiliation{School of Physics, Dalian University of Technology, 116024 Dalian, China}

\begin{abstract}
We present a deperturbation analysis of the spin-orbit coupled $\rm A^1\Sigma^+$ and $\rm b^3\Pi_{0^+}$ states of LiRb based on the rovibrational energy levels observed previously by photoassociation spectroscopy in bosonic $^7$Li$^{85}$Rb molecule. Using the genetic algorithm, we fit the potential energy curves of the $\rm A^1\Sigma^+$ state and the $\rm b^3\Pi$ state into point-wise form. We then fit these point-wise potentials along with the spin-orbit coupling into expanded Morse oscillator functional form and optimise analytical parameters based on the experimental data. From the fitted results, we calculate the transition dipole moment matrix elements for transitions from the rovibrational levels of the coupled $\rm A^1\Sigma^+$-$\rm b^3\Pi_{0^+}$ state to the Feshbach state and the absolute rovibrational ground state for fermionic $^6$Li$^{87}$Rb molecule. Based on the calculated transition dipole moment matrix elements, several levels of the coupled $\rm A^1\Sigma^+$-$\rm b^3\Pi_{0^+}$ state are predicted to be suitable as the intermediate state for stimulated Raman adiabatic passage transfer from the Feshbach state to the absolute rovibrational ground state. In addition, we also provide a similar estimation for ${\rm B}^1\Pi$-${\rm c}^3\Sigma_1^+$-${\rm b}^3\Pi_1$ state based on available $ab\ initio$ interaction potentials.
\end{abstract}
\maketitle
\section{\label{sec1}Introduction}
Since the turn of the century, the realm of ultracold atoms and molecules has emerged as one of the most fascinating fields for physicists \cite{BookHuiZhai2021}  and chemists \cite{BookJesusRios2020}. Compared to ultracold atoms, ultracold molecules offer a richer internal structure, enhancing application capabilities. The most common techniques used to prepare ultracold molecules in experiments are photoassociation (PA) \cite{Rev1PA2006,Rev2PA12012} and magnetoassociation (MA) \cite{BookColdMol2009}. The latter technique prepares Feshbach molecules \cite{RevFB2010} that are subsequently transferred to their absolute rovibrational ground state via stimulated Raman adiabatic passage (STIRAP) \cite{RevSTIRAP2017}. It has been productively adopted on bialkali polar molecules including $^{6}$Li$^{23}$Na \cite{MA6Li23Na2017}, $^{6}$Li$^{40}$K \cite{MALiK2024reprint}, $^{23}$Na$^{39}$K \cite{MANa23K392020}, $^{23}$Na$^{40}$K \cite{MANa23K402015,MANa23K402018,MANa23K402019,MANa23K402021,MANa23K402023}, $^{23}$Na$^{87}$Rb \cite{MA1Na23Rb872016,MA2Na23Rb872017}, $^{23}$Na$^{133}$Cs \cite{MANaCsNi2023,MANaCsStevenson2023}, $^{40}$K$^{87}$Rb \cite{MAK40Rb872008}, and $^{87}$Rb$^{133}$Cs \cite{MARbCsCornish2014,MARb87Cs133Naegerl2014,MARb87Cs1332023}. 

However, ultracold LiRb molecules, especially the fermionic species, have not been deterministically prepared in the absolute rovibrational ground state via MA and subsequently STIRAP. LiRb molecule has several unique features in comparison to the species mentioned above \cite{ICSThesis2018,BlasingThesis2018}. It exhibits the second largest body-frame dipole moment among fermionic bialkali molecules \cite{ZiangPRA2020}. Microwave-shielded LiRb molecules promise orders of magnitude higher good-to-bad collision rate ratios compared to existing fermionic polar molecules \cite{ValtolinaNature2020, SchindewolfNature2022}. As a result, they have the potential to be cooled to unprecedentedly low temperatures for the realization of exotic quantum matters \cite{TaoShiPRL2022}. This is made possible by their large dipole moment and high rotational constant. 

In experiment, LiRb molecule has been investigated through spectroscopic techniques. Earlier in 2009, Marzok $et\ al.$ observed Feshbach resonance \cite{FBZhangrong2022} in an ultracold $^7$Li and $^{87}$Rb mixture \cite{FBLi7Rb872009}. Ivanova $et\ al.$ observed spectroscopic data of the $\rm X^1\Sigma^+$, $\rm a^3\Sigma^+$, $\rm B^1\Pi$ and $\rm D^1\Pi$ states \cite{TiemannIPAX1Sa3S2011,TiemannLEVELB1PiD1Pi2013}. Dutta $et\ al.$ studied the first 21 vibrational levels of the B$^1\Pi$ via laser-induced fluorescence spectroscopy \cite{LiRbLEVEL1B1Pi2011}, and firstly produced ultracold bosonic $^7$Li$^{85}$Rb molecules in electronic excited states by PA \cite{add1Dutta2013,add2Dutta2014}. In 2014, Lorenz $et\ al.$ observed the transitions from the ground electronic state to the $\rm B^1\Pi$ and $\rm D^1\Pi$ states via PA and resonantly enhanced multiphoton ionisation (REMPI) spectroscopy \cite{LiRbLEVEL3B1PiD1Pi2014}. In 2015, Altar $et\ al.$ probed and identified the high-lying vibrational levels of $\rm a^3\Sigma^+$ state and the low-lying vibrational levels of $(3)^3\Pi$ and $(4)^3\Sigma^+$ \cite{LiRbLEVELa3Sig+f3Pig3Sig+2015}. In 2016, Blasing $et\ al.$ investigated the low-lying vibrational levels of $\rm a^3\Sigma^+$ state \cite{PALiRba3Sig+2016}. Stevenson $et\ al.$ explored the rovibrational levels in the ${\rm A}^1\Sigma^+$, ${\rm b}^3\Pi$, $\rm C^1\Sigma^+$ and $\rm d^3\Pi$ states \cite{LiRbLEVELA1b3C12016,LiRbLEVELd3Pi2016}, and detected ultracold absolute rovibrational ground-state $^7$Li$^{85}$Rb molecules produced through PA followed by spontaneous decay via REMPI \cite{PALi7Rb852016}. In 2017, Dutta $et\ al.$ carried out two-photon PA experiments and proposed a STIRAP pathway to directly convert free $^7$Li and $^{85}$Rb atoms to weakly bound triplet $^{7}$Li$^{85}$Rb molecules \cite{add3Dutta2017}. 

In addition to experimental studies, there have been many theoretical studies on the underlying interactions in LiRb. Earlier in 2009, Korek $et\ al.$ calculated the potential energy curves (PECs) for the 58 lowest electronic states of the LiRb molecule using the configuration interaction method, and the spin-orbit (SO) effect was considered through a semi-empirical SO pseudo-potential \cite{abPECLB2009}. In 2012, Jendoubi $et\ al.$ calculated the adiabatic and diabatic PECs, the permanent dipole moments and the transition dipole moments (TDMs) for 38 electronic states of LiRb \cite{abTDM2012}. In 2016, You $et\ al.$ investigated the spin-forbidden cooling of LiRb molecule and calculated the PECs of the $\rm X^1\Sigma^+$, $\rm a^3\Sigma^+$, $\rm b^3\Pi$ and $\rm B^1\Pi$ states, as well as the SO coupling effects and TDMs using the multi-reference configuration interaction (MRCI) method \cite{abPECTDMCN2016}.  Employing the quasi-relativistic electronic wave functions obtained by the MRCI method, Bormotova $et\ al.$ calculated TDMs and SO coupling matrix elements between the ground and excited states for LiRb \cite{abTDMPCCP2018,abSOCPRA2019}. In 2020, Kozlov $et\ al.$ evaluated PECs and SO coupling matrix elements in low-lying electronic states of LiRb molecules using two different approaches \cite{abPECSOCPCCP2020}. 

Although the spectroscopic data of the ${\rm A}^1\Sigma^+$ and ${\rm b}^3\Pi_{0^+}$ states for $^7$Li$^{85}$Rb have been observed and assigned in \cite{LiRbLEVELA1b3C12016}, deperturbation analysis \cite{BookSpec&Dyna2004} has not been adopted on these electronic states of LiRb. In the past few years, deperturbation analysis has already been adopted on the ${\rm A}^1\Sigma^+$ and ${\rm b}^3\Pi_{0^+}$ states for NaRb \cite{DP0NaRb2002,DPNaRb2007}, RbCs \cite{DP0RbCs2003,DP1RbCs2010,DP2RbCs2014}, Na$_2$ \cite{DPNaNa2007}, Cs$_2$ \cite{DP1CsCs2011,DP2CsCs2019}, NaCs \cite{DPNaCs2009}, KCs \cite{DP1KCs2010,DP2KCs2010}, LiCs \cite{DPLiCs2015}, NaK \cite{DPNaK2015}, and KRb \cite{DPKRb2016}. Besides experimental and theoretical data, efficient optimisation algorithms are essential for deperturbation analysis. One of the most commonly used algorithms is a modified Levenberg-Marquardt (LM) algorithm \cite{BookNumRec2007}. Following the surge in computational power, heuristic algorithms like genetic algorithm (GA) have been applied to optimisation tasks in atomic and molecular physics, such as fitting the single-channel PEC based on spectroscopic data \cite{GANaLi2008,GARbCs2011,GALiRb2019} and optimising parameters of the laser pulse \cite{GALD2023}. Inspired by these
works, we present a genetics-based deperturbation analysis to fit the PECs of ${\rm A}^1\Sigma^+$ and ${\rm b}^3\Pi$ states and the SO coupling between them based on the spectroscopic data in \cite{LiRbLEVELA1b3C12016}. With the fitted PECs and SO coupling, we predict the rovibrational levels of the coupled ${\rm A}^1\Sigma^+$ and ${\rm b}^3\Pi_{0^+}$ states which are suitable as the intermediate state for STIRAP transfer from the Feshbach state to the absolute rovibrational ground state. This is vital for preparing the ultracold LiRb molecule in experiments. 

The remaining part of this paper proceeds as follows: the second section introduces the methodology used in the present work; the third section discusses the fitted results and future prospects; the last section states the conclusion of the present work. 

\section{\label{Sec2}Methodologies}
\subsection{\label{Sec2A}model Hamiltonian}

\begin{figure}[tbp]
	\centering
	\includegraphics[width=0.45\textwidth]{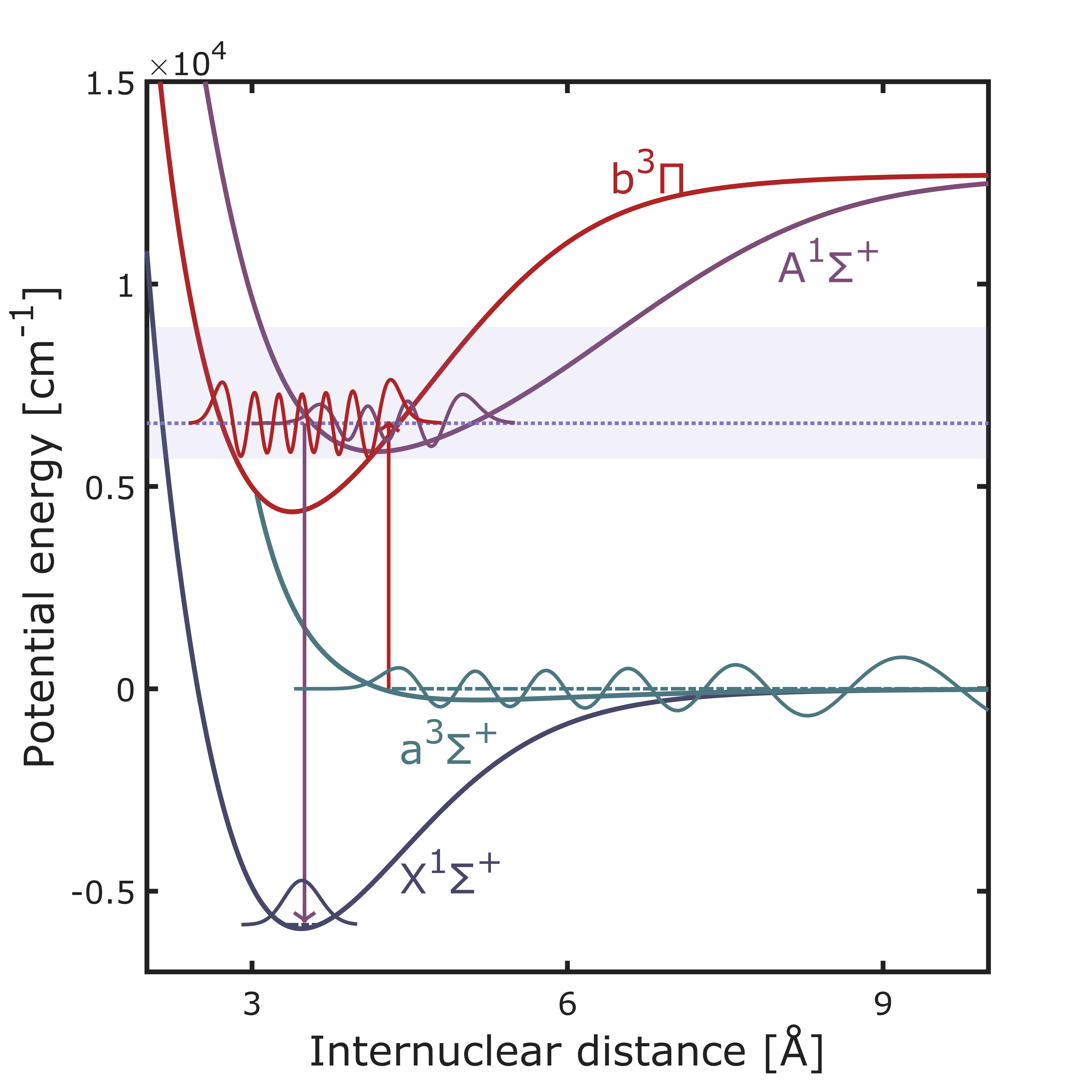}
	\caption{ The low-lying electronic states of the LiRb molecule. The PECs of the $\rm X^1\Sigma^+$ and $\rm a^3\Sigma^+$ states are from \cite{TiemannIPAX1Sa3S2011}. The PECs of the $\rm A^1\Sigma^+$ and $\rm b^3\Pi$ states are from \cite{abPECSOCPCCP2020}. The shaded area denotes the observed energy range in \cite{LiRbLEVELA1b3C12016}, which is from 5682 $\rm cm^{-1}$ to 8943 $\rm cm^{-1}$. The green and dark purple wave functions are for the Feshbach state and the absolute rovibrational ground state. The red and light purple wave functions are for the intermediate state chosen from the SO coupled $\rm A^1\Sigma^+$ and $\rm b^3\Pi$ states. The red and light purple arrows indicate the transitions induced by the pump and Stokes lasers, respectively.}
	\label{abPECsFig}
\end{figure}
The $\rm A^1\Sigma^+$ and $\rm b^3\Pi$ states are two of the lowest electronic excited states for bialkali polar molecules. Together with $\rm X^1\Sigma^+$ and $\rm a^3\Sigma^+$ states, the PECs of $\rm A^1\Sigma^+$ and $\rm b^3\Pi$ of LiRb are illustrated in Fig. \ref{abPECsFig}. In order to establish an effective model Hamiltonian for the SO coupled $\rm A^1\Sigma^+$ and $\rm b^3\Pi$ states, one should typically consider the diabatic PECs of $\rm A^1\Sigma^+$ and $\rm b^3\Pi$ states and the SO splitting of $\rm b^3\Pi$ state, as well as the SO coupling between the $\rm A^1\Sigma^+$ and $\rm b^3\Pi_{0^+}$ states and the spin-rotational interaction between the $\Omega=0^+,\:1,\:2$ components of $\rm b^3\Pi$ state. However, the spectroscopic data in \cite{LiRbLEVELA1b3C12016} are limited to the $\Omega=0^+$ components of the $\rm b^3\Pi$ states, and there are no data from the $\Omega=1, 2$ components. Therefore, we consider only $\rm A^1\Sigma^+$ and $\rm b^3\Pi_{0^+}$ states in the present work.

The relevant coupled-channel (CC) radial Schrödinger equation is
\begin{equation}\label{eqH}
\left(\begin{array}{ll}T+V_{\rm A} & V_{\rm Ab}\\V_{\rm Ab} & T+V_{\rm {b0}}\\\end{array}\right)
\left(\begin{array}{ll}\Phi_{\rm A} \\\Phi_{\rm b0}\\\end{array}\right)
=E^{\rm CC}_{v'}
\left(\begin{array}{ll}\Phi_{\rm A} \\\Phi_{\rm b0}\\\end{array}\right),
\end{equation}
where  $T$ is the kinetic energy term, $V_{\rm A}$, $V_{\rm b0}$ are the potential energy terms, $V_{\rm Ab}$ is the SO coupling term, $v'$ indexes the rovibrational levels of coupled ${\rm A}^1\Sigma^+$-${\rm b}^3\Pi_{0^+}$ state, $E_{v'}^{\rm CC}$ is its rovibrational energy, and $\Phi_{\rm A}$ and $\Phi_{\rm b0}$ are two components of the eigenstate. The fractional partitions of ${\rm A}^1\Sigma^+$ state and ${\rm b}^3\Pi_{0^+}$ state in the eigenstate are $P_{\rm A}=	\left\langle\Phi_{\rm A}|\Phi_{\rm A}\right\rangle$ and $P_{\rm b0}=\left\langle\Phi_{\rm b0}|\Phi_{\rm b0}\right\rangle$. The eigenstate is assigned to the electronic state with larger fractional partition. 

The potential energy and SO coupling terms in Eq. \eqref{eqH} are
\begin{equation}\label{eqV}
\begin{array}{lll}V_{\rm A}=U_{\rm A}+E^{\rm ROT}_{\rm A},  \\
	V_{\rm b0}=U_{\rm b}-\xi_{\rm bb}+E^{\rm ROT}_{\rm b0},\\
	V_{\rm Ab}=-\sqrt2\xi_{\rm Ab},\\
	
\end{array}
\end{equation}
where $U_{\rm A}$ and $U_{\rm b}$ stand for the diabatic PECs of $\rm A^1\Sigma^+$ state and $\rm b^3\Pi$ state, respectively, $\xi_{\rm bb}$ is the SO splitting of the $\rm b^3\Pi$ state, $\xi_{\rm Ab}$ is the SO coupling between the $\rm A^1\Sigma^+$ state and $\rm b^3\Pi_{0^+}$ state, and $E^{\rm ROT}_{\rm A}$ and $E^{\rm ROT}_{\rm b0}$ are the rotational Hamiltonians of the $\rm A^1\Sigma^+$ and $\rm b^3\Pi_{0^+}$ states, respectively. Notably, the SO coupling term $\xi_{\rm Ab}$ in $V_{\rm Ab}$ is multiplied by a sign-consistent multiplicative factor of $-\sqrt{2}$, which corrects the fine splitting near the dissociation thresholds \cite{abSOCPRA2019}. The rotational Hamiltonians \cite{BookSpec&Dyna2004} in Eq. \eqref{eqV} are derived by
\begin{widetext}
\begin{equation}\label{eqROT1}
	E^{\rm ROT}_{^{2S+1}\Lambda_{\Omega}}=\frac{\hbar^2}{2\mu R^2}[J(J+1)-\Omega^2+S(S+1)-\Sigma^2+L(L+1)-\Lambda^2],\\
\end{equation}
\end{widetext}
where $\hbar$ is the reduced Planck constant, $\mu$ is the reduced mass, $R$ is the internuclear distance, $S$ is the total electron spin, $L$ is the total electronic orbital angular momentum, $J$ is the total molecular angular momentum exclusive of nuclear spin, $\Sigma$ and $\Lambda$ are the projection of $S$ and $L$ along the internuclear axis, and $\Omega=|\Sigma+\Lambda|$. Therefore, \begin{equation}\label{eqROT2}
	\begin{array}{lll}
		E^{\rm ROT}_{\rm A}=E^{\rm ROT}_{\rm b0}=\frac{\hbar^2}{2\mu R^2}\left[J(J+1)+2\right].\\
	\end{array}
\end{equation}

In the present work, the higher-order SO terms and hyperfine coupling are ignored due to the limitation of spectroscopic data. Within the framework of our model Hamiltonian, the terms $U_{\rm A}$, $U_{\rm b}$, and $\xi_{\rm Ab}$ are supposed to be optimised. The optimisation algorithm is detailed in the next two subsections. 

\subsection{\label{Sec2B}Genetic algorithm}
\begin{figure}[htbp]
	\centering
	\includegraphics[width=0.43\textwidth]{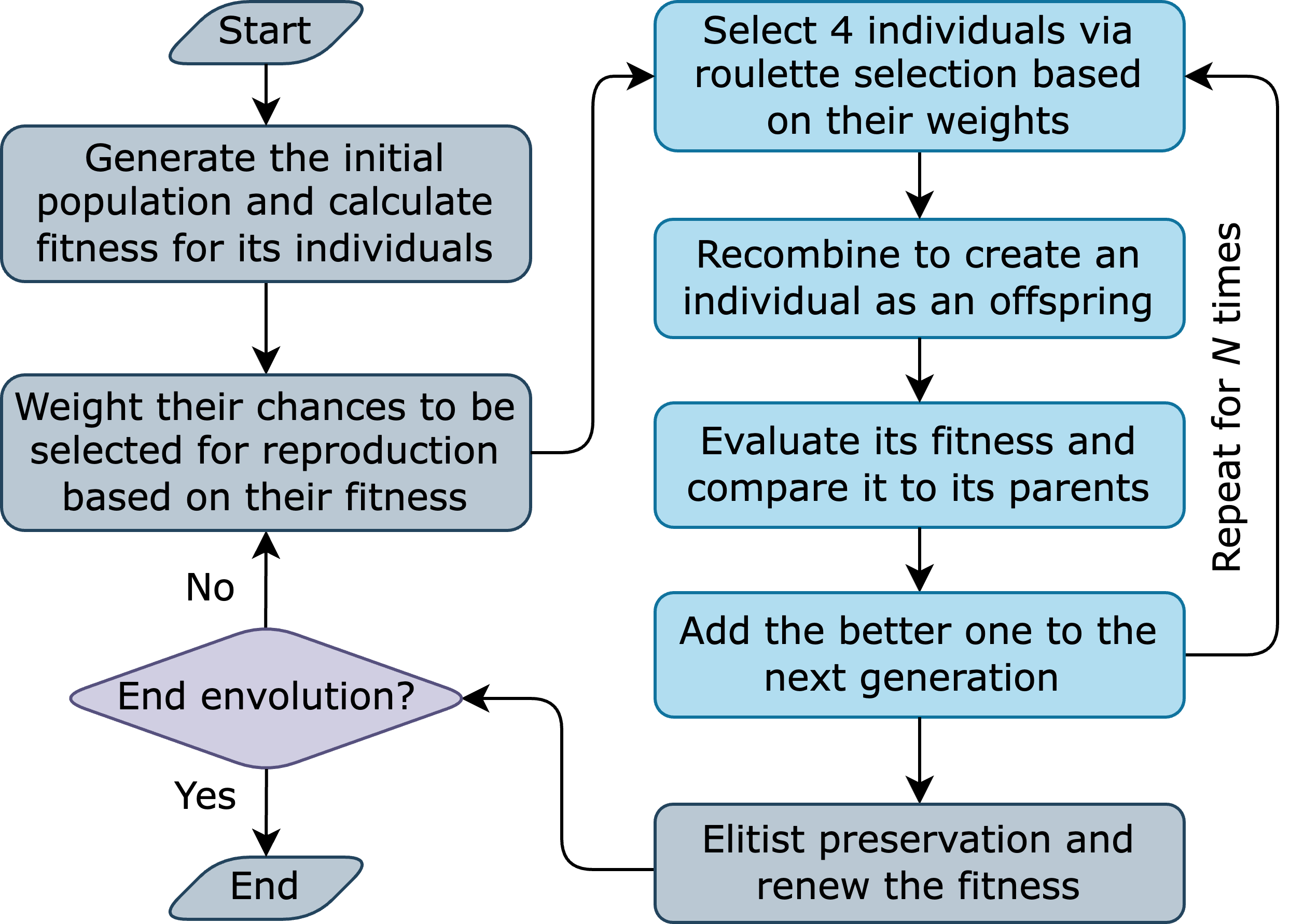}
	\caption{The flowchart of the genetic algorithm adopted in this work.}
	\label{flowchartFig}
\end{figure}

GA is a stochastic global optimising method inspired mainly by natural selection and genetics \cite{RevGA2021}. In the present work, we modify the algorithm developed in \cite{GALiRb2019} to optimise the point-wise $U_{\rm A}$ and $U_{\rm b}$, which has been previously calculated in \cite{abPECSOCPCCP2020} through the configuration interaction
method with core-valence correlation accounted for using core polarization potentials (CI-CPP). The flowchart of our modified algorithm is illustrated in Fig. \ref{flowchartFig}, and each part of it is elaborated in the following. 

\subsubsection{\label{Sec2B1}Initial population}
\begin{figure}[htbp]
	\centering
	\includegraphics[width=0.4\textwidth]{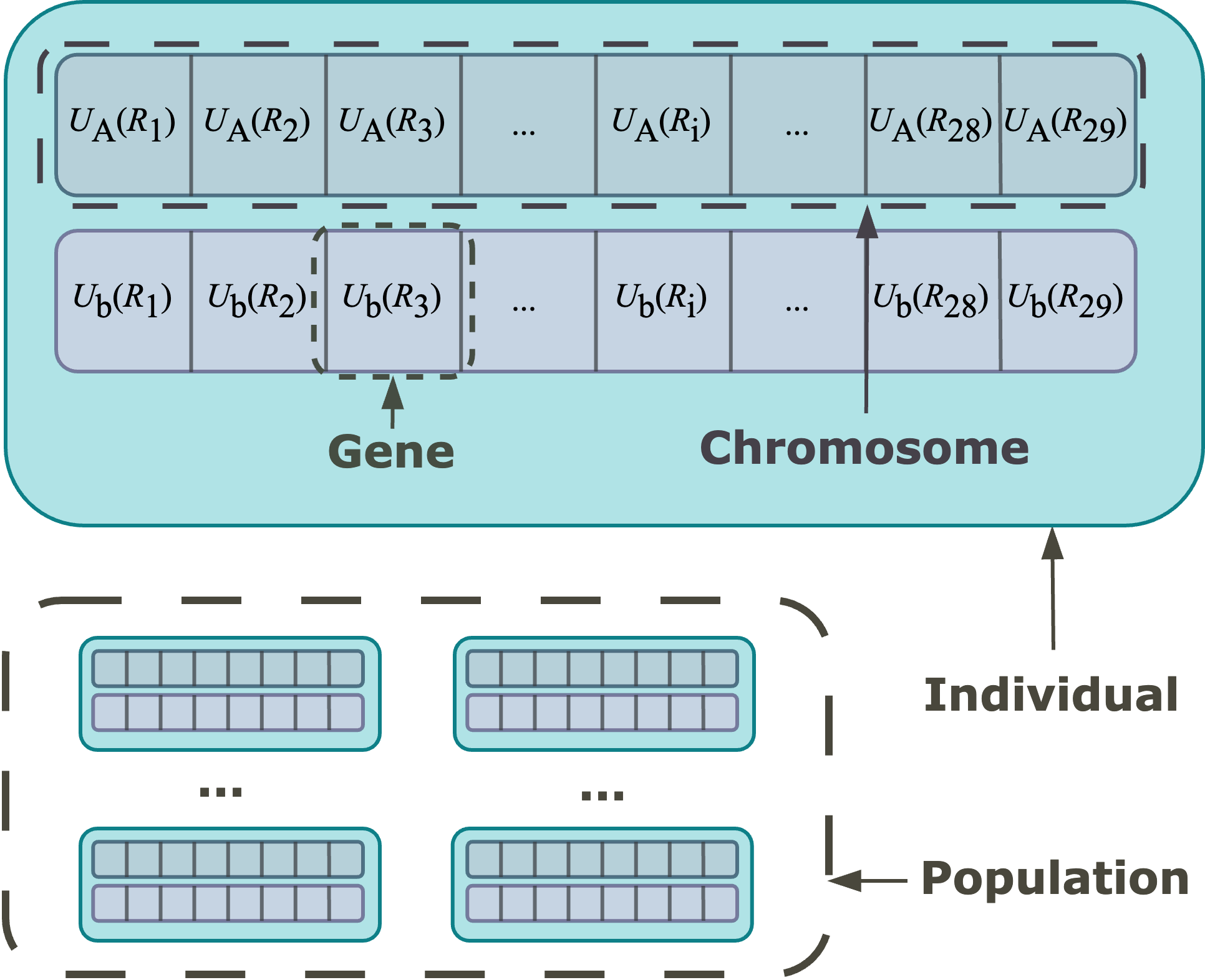}
	\caption{The schematic diagram of a population in the genetic algorithm of this work.}
	\label{popMap}
\end{figure}

Our algorithm starts with creating an initial population that contains $N$ individuals, each representing a candidate solution to the current problem. There are two chromosomes in each individual, which represent the diabatic PECs $U_{\rm A}$ and $U_{\rm b}$, respectively. Both of them are point-wise PECs along the non-uniform $R$-space grid, as listed in the first column in Table  \ref{tableGAPEC}. There are 29 genes in each chromosome, each of which stands for a potential point. A schematic diagram of the population is shown in Fig. \ref{popMap}. 
\begin{table}[htbp]
	\caption{\label{tableGAPEC}
		The point-wise PECs of the $\rm A^1\Sigma^+$ and $\rm b^3\Pi $ states of LiRb obtained via $ab\ initio$ calculations in \cite{abPECSOCPCCP2020} ($U_{\rm A}^{ab}$ and $U_{\rm b}^{ab}$) and GA optimisation ($U_{\rm A}^{\rm GA}$ and $U_{\rm b}^{\rm GA}$). Internuclear distances are given in $\rm\AA$, and energies are in $\rm cm^{-1}$.}
	\begin{ruledtabular}
		\begin{tabular}{rrrrr}
			$R$  & $U_{\rm A}^{ab}$ & $U_{\rm b}^{ab}$ & $U_{\rm A}^{\rm GA}$ & $U_{\rm b}^{\rm GA}$ \\ \hline
			2.0  & 26891.94 & 18627.56 & 26812.65 & 18696.66 \\
			2.2  & 21646.73 & 13344.37 & 21720.34 & 13399.84 \\
			2.4  & 17618.69 & 9853.42  & 17597.78 & 9173.52  \\
			2.6  & 14333.90 & 7512.10  & 14320.65 & 6988.95  \\
			2.8  & 11691.20 & 5954.89  & 11768.61 & 5625.15  \\
			3.0  & 9660.52  & 4992.17  & 9828.51  & 4734.67  \\
			3.2  & 8180.83  & 4500.57  & 8134.73  & 4247.30  \\
			3.4  & 7157.70  & 4374.61  & 7089.34  & 4123.49  \\
			3.6  & 6493.40  & 4523.65  & 6419.08  & 4281.63  \\
			3.8  & 6101.02  & 4872.14  & 6010.71  & 4655.91  \\
			4.0  & 5908.97  & 5359.15  & 5803.91  & 5183.73  \\
			4.2  & 5861.14  & 5936.55  & 5750.32  & 5808.80  \\
			4.4  & 5915.47  & 6566.73  & 5811.50  & 6483.36  \\
			4.6  & 6042.22  & 7220.42  & 5945.16  & 7169.24  \\
			4.8  & 6221.29  & 7875.21  & 6127.08  & 7837.83  \\
			5.0  & 6443.57  & 8515.00  & 6357.31  & 8472.52  \\
			5.4  & 6983.73  & 9680.68  & 6919.18  & 9619.65  \\
			5.8  & 7624.96  & 10639.51 & 7587.32  & 10665.27 \\
			6.2  & 8333.23  & 11357.22 & 8296.31  & 11412.43 \\
			6.6  & 9063.86  & 11846.76 & 9029.58  & 11911.12 \\
			7.0  & 9775.01  & 12159.70 & 9812.40  & 12159.70 \\
			8.0  & 11260.91 & 12520.23 & 11295.72 & 12520.23 \\
			9.0  & 12125.39 & 12641.24 & 12107.77 & 12641.24 \\
			10.0 & 12486.87 & 12688.89 & 12486.87 & 12688.89 \\
			11.0 & 12622.40 & 12711.20 & 12622.40 & 12711.20 \\
			12.0 & 12678.41 & 12722.59 & 12678.41 & 12722.59 \\
			14.0 & 12704.88 & 12729.11 & 12704.88 & 12729.11 \\
			17.0 & 12718.63 & 12733.00 & 12718.63 & 12733.00 \\
			21.0 & 12726.15 & 12735.30 & 12726.15 & 12735.30   \\
		\end{tabular}
	\end{ruledtabular}
\end{table}

To generate the chromosomes and ensure the physical characteristics of the point-wise PECs, Wei Hua's (HW) 4-parameter potential function \cite{WeiPOT1990} is adopted, which is expressed as
\begin{equation}\label{eqHW}
U_{\rm HW}(R)=T_e+D_e\left[\frac{1-e^{-b\cdot(R-R_e)}}{1-Ce^{-b\cdot(R-R_e)}}\right]^2,
\end{equation}
where $T_e$ is the potential energy minimum, $D_e$ is the potential well depth, $R_e$ is the equilibrium internuclear distance, and $b$ and $C$ are two parameters controlling the shape of the potential well. $T_e + D_e$ is the dissociation threshold of Li 2S + Rb 5P, which is fixed to 12737.35 $\rm cm^{-1}$ according to the NIST atomic database \cite{NISTDatabase}. Notably, the Li 2S + Rb 5S atomic asymptote serves as the benchmark for energies. 

\begin{figure}[htbp]
	\centering
	\includegraphics[width=0.45\textwidth]{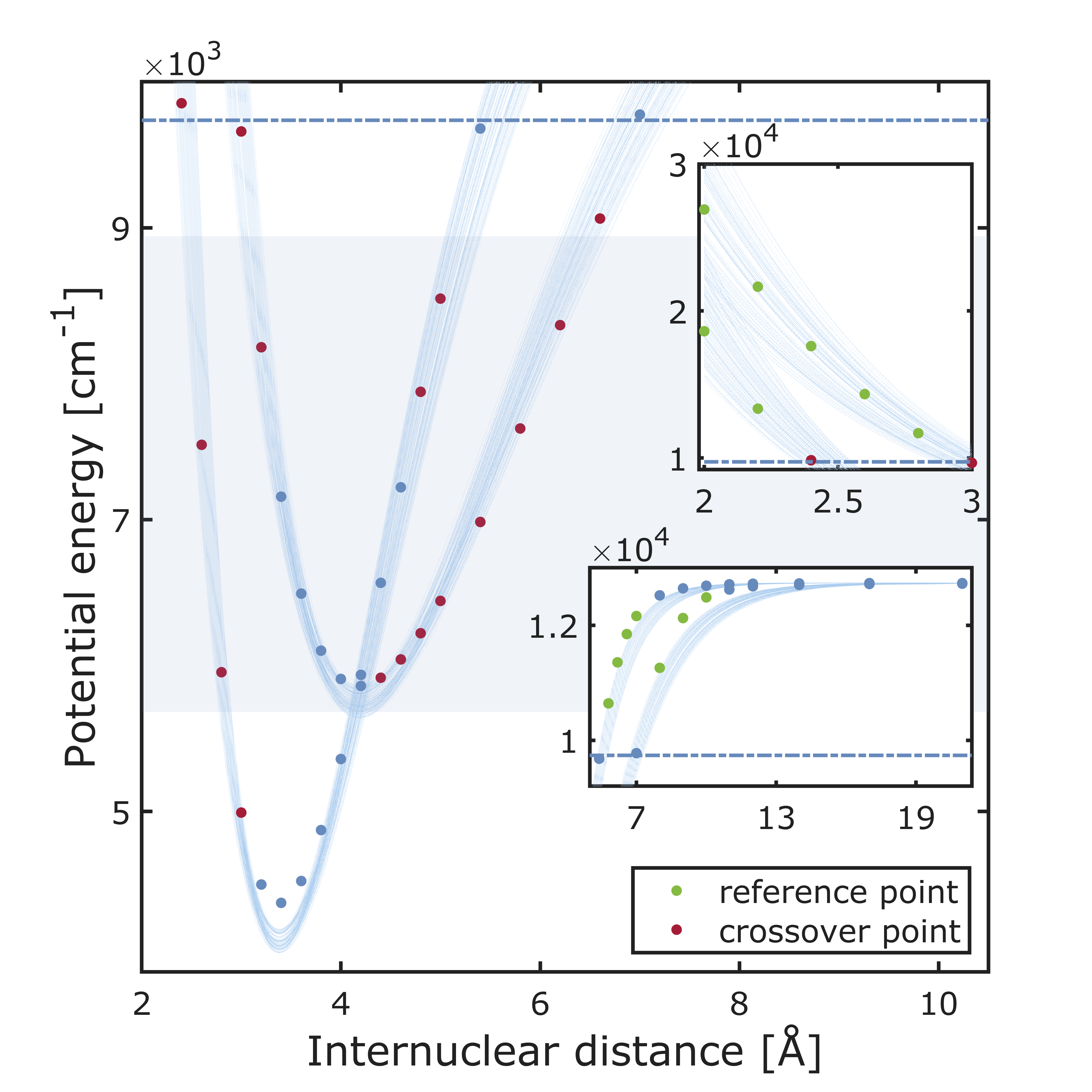}
	\caption{The diagram of encoding point-wise PECs. The subfigures illustrate potential points in the short-range and long-range regions. The cluster of light blue lines denotes HW potential with different parameters. The colored points are the $ab\ initio$ potential points listed in Table \ref{tableGAPEC}. The green points are the reference points mentioned in Appendix \ref{App1}. The red points could be chosen as crossover point, as mentioned in Section \ref{Sec2B3}. The shaded area denotes the energy range of the observed rovibrational levels. The dot-dashed line denotes the upper energy boundary of the reference levels in Appendix \ref{App1}. }
	\label{iniPop}
\end{figure}

To create the individuals in the initial population, 4 parameters $T_e$, $R_e$, $b$ and $C$ are tuned, each of which is uniformly distributed around a central value with a range. The determination of the central values and ranges is discussed below. The potential minimum $T_e$ is estimated in \cite{LiRbLEVELA1b3C12016} by fitting the spectroscopic data to 
\begin{equation}\label{eqTv}
T_v=T_e+\omega_e(v+\frac{1}{2})-\omega_e\chi_e(v+\frac{1}{2})^2+\omega_ey_e(v+\frac{1}{2})^3,
\end{equation}
where $\omega_e$ is the vibrational harmonic frequency, $\omega_e\chi_e$ and $\omega_ey_e$ are first and second anharmonic correction coefficients, $v$ is the vibrational quantum number of the uncoupled ${\rm A}^1\Sigma^+$ or ${\rm b}^3\Pi_{0^+}$ states, and $T_v$ is the rovibrational energy. The parameters $b$ and $C$ are estimated by fitting the $ab\ initio$ PECs in \cite{abPECSOCPCCP2020} with HW potential function. The equilibrium distance $R_e$ is set to the value theoretically calculated in \cite{abPECSOCPCCP2020}. Their central values and ranges are listed in Table  \ref{tableHW}. As seen, different to $T_e$, $b$ and $C$, all individuals share the same $R_e$ in the initial population. That is because the PA spectroscopy data in \cite{LiRbLEVELA1b3C12016} is limited to the rotational level of $J=1$, the value of $R_e$ related to the rotational constant cannot be fitted from the observed levels. The point-wise $U_{\rm A}$ and $U_{\rm b}$ corresponding to different $T_e$, $R_e$, $b$ and $C$, as illustrated in Fig. \ref{iniPop}, are encoded into the population. 
\begin{table}[htbp]
	\caption{\label{tableHW}
		The central values and ranges of HW parameters for the $\rm A^1\Sigma^+$ and $\rm b^3\Pi $ states of LiRb. The second and third columns are the central values, and the forth column denotes their ranges. The range is from $1-\Delta$ to $1+\Delta$ multiplied by the central value. Energies are given in $\rm cm^{-1}$, internuclear distances are in $\rm\AA$, coefficient $b$ is in the uint of 1/$\rm\AA$, and $C$ is dimensionless.}
	\begin{ruledtabular}
		\begin{tabular}{rrrr}
			& $U_{\rm A}$ & $U_{\rm b}$   & $\Delta$ \\ \hline
			$T_e$     & 5756.6        & 4112.6     & 2\%      \\
			$T_e+D_e$ & 12737.3467    & 12737.3467 & 0        \\
			$R_e$     & 4.1836        & 3.3802     & 0        \\
			$b$         & 0.5269        & 0.8309     & 10\%     \\
			$C$         & -0.0827       & -0.1979    & 10\%    
		\end{tabular}
	\end{ruledtabular}
\end{table}

\subsubsection{\label{Sec2B2}Fitness function}
After the initial population is created, it naturally follows to define a fitness function that evaluates the performance of individuals. In order to remedy the limitation of observed energy range in \cite{LiRbLEVELA1b3C12016}, as shown in Fig. \ref{abPECsFig} and \ref{iniPop}, the fitness function in our algorithm should take not only rovibrational levels but also potential points into account. It contains two terms denoted as $F_\alpha$ and $F_\beta$, concerning potential points and rovibrational levels, respectively, as described in Appendix \ref{App1}.

\begin{figure}[htbp]
	\centering
	\includegraphics[width=0.45\textwidth]{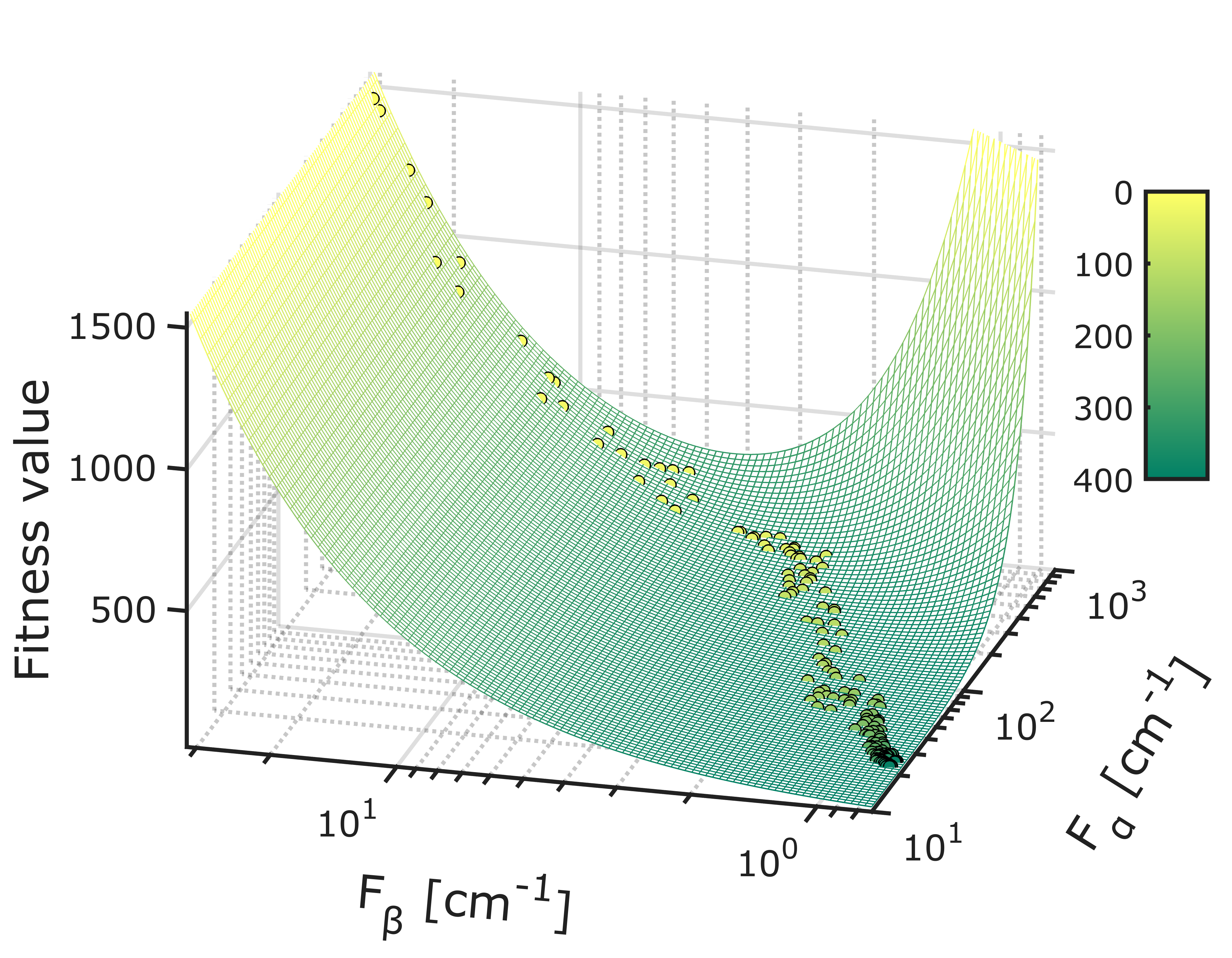}
	\caption{The variation of $F$ versus $F_\alpha$ and $F_\beta$ with $\omega_{F}=0.002$. The fittest individuals in each generation are denoted by circled dots, while the colorbar signifies generations. }
	\label{fnSurfFig1}
\end{figure}

Considering the difference in the order of magnitude between $F_\alpha$ and $F_\beta$, the fitness function in our algorithm is expressed as
\begin{equation}\label{eqF}
F=\frac{F_\beta}{\sigma_\beta}+\omega_{F}\cdot\left(\frac{F_\alpha}{F_\beta}\right)^2,
\end{equation}
where $\sigma_\beta$ is the experimental uncertainty and $\omega_{F}$ is an coefficient merging $F_{\alpha}$ and $F_{\beta}$. It needs to set $\omega_{F}$ properly to control the evolution. At the beginning of evolution, the fitting should primarily focuses on the rovibrational levels. As the evolution goes, the potential points are also supposed to be fitted together with the rovibrational levels. The variation of the fitness value $F$ versus $F_\alpha$ and $F_\beta$ is shown in Fig. \ref{fnSurfFig1}. The individuals with smallest fitness value in the current population, namely the fittest individuals, are denoted by the circled dots for each generation.

The smaller the fitness value of an individual, the better it is. To select better individuals and speed up the convergence, the weight for each individual is assigned based on their fitness, as
\begin{equation}\label{eqW}
W_n=\frac{\left(\frac{\bar F}{F_n}\right)^\frac{1}{2}}{\sum_n\left(\frac{\bar F}{F_n}\right)^\frac{1}{2}},
\end{equation}
where $n$ indexes individuals, $F_n$ is the fitness value of an individual, $\bar F$ is the median fitness value among all individuals, the exponential term of $\frac{1}{2}$ is to reduce the effect of extreme individuals \cite{ICSThesis2018}. The weight of an individual stands for its probability of being selected via roulette selection. 

\subsubsection{\label{Sec2B3}Evolutionary operation}
The evolutionary operation for GA generally includes selection, crossover and mutation \cite{GAYin2024}. In the present work, the evolutionary operation is adapted from the methods in \cite{GALiRb2019}, which works as follows: 

\begin{enumerate}
\item Select 4 individuals from the population via roulette selection and randomly specify a chromosome to evolve. The chromosomes specified to evolve in these 4 individuals are labelled $\boldsymbol{\mathit{x}}$, $\boldsymbol{\mathit{y}}$, $\boldsymbol{\mathit{z}}$, $\boldsymbol{\mathit{w}}$, respectively. 

\item Create a new chromosome $\boldsymbol{\mathit{u}}$ according to
\begin{equation}\label{eqx1}
\boldsymbol{\mathit{u}}=\boldsymbol{\mathit{x}}+\lambda(\boldsymbol{\mathit{y}}-\boldsymbol{\mathit{z}}),
\end{equation}
where $\lambda$ is a hyperparameter related to the evolutionary intensity.

\item Choose a crossover point \footnote[1]{Since the energy range of the observed rovibrational levels is limited, the crossover points should also be limited in certain regions}. The crossover point is randomly chosen from the red points shown in Fig. \ref{iniPop}.

\item Cross $\boldsymbol{\mathit{u}}$ with $\boldsymbol{\mathit{w}}$ according to
\begin{equation}\label{eqx2}
v_i=
\left\{\begin{array}{lc}
	w_i&{\rm if}\enspace i\leq I 
	\\u_i&{\rm if}\enspace i>I
\end{array} \right.,
\end{equation}
when the crossover point is to the left of the equilibrium distance, otherwise, 
\begin{equation}\label{eqx3}
v_i=
\left\{\begin{array}{lc}
	 u_i& {\rm if}\enspace i\leq I
	 \\w_i&{\rm if}\enspace i>I
\end{array}\right.,
\end{equation}
where $i$ indexes genes and $I$ is the index of the crossover point. The crossover probability is $P_{\rm x}$. If the crossover doesn't occur, set $\boldsymbol{\mathit{v}}=\boldsymbol{\mathit{u}}$ and proceed directly to the next step. 

\item Assemble $\boldsymbol{\mathit{v}}$ with the unevolved chromosome in the old individual containing $\boldsymbol{\mathit{w}}$. Compare the fitness of this new individual to the old individual containing $\boldsymbol{\mathit{w}}$, and add the better one to the next generation.
\end{enumerate}

The steps above is repeated $N$ times to obtain the next generation population. After that, elitist preservation is adopted to ensure global convergence, which directly copies the best individual emerged so far and replaces the current worst individual \cite{GAYin2024}. The above operation is iterated to evolve the population until the $T$-th generation, and the fitted point-wise PECs $U_{\rm A}$ and $U_{\rm b}$ is obtained.
\subsection{\label{Sec2C}Analytical potential functions}
After the point-wise $U_{\rm A}$ and $U_{\rm b}$ are optimised via GA, we adopt an optimisation algorithm called sequential least squares programming (SLSQP) \cite{SLSQP1988} to further optimise $U_{\rm A}$, $U_{\rm b}$ and $\xi_{\rm Ab}$ in the expanded Morse oscillator (EMO) function form. The EMO function is expressed as
\begin{equation}\label{eqEMO}
U_{\rm EMO}(R)=T_e+D_e[1-e^{-\beta_{\rm EMO}(R)\cdot(R-R_e)}]^2,
\end{equation}
and the coefficient $\beta_{\rm EMO}$ is a polynomial
\begin{equation}\label{eqB}
\beta_{\rm EMO}(R)=\sum_{i=0}^{N_\beta}\beta_i\left(\frac{R^q-R_{\rm ref}^q}{R^q+R_{\rm ref}^q}\right)^i,
\end{equation}
where $q$ is a small positive integer, and $R_{\rm ref}$ is a positive reference distance slightly larger than $R_e$, both of which are fixed. 

Firstly, the point-wise $U_{\rm A}$ and $U_{\rm b}$ obtained by GA and $\xi_{\rm Ab}$ obtained by $ab\ initio$ calculations in \cite{abPECSOCPCCP2020} are fitted into EMO form through betaFIT program \cite{betaFIT2017}. The parameters fitted via betaFIT serve as initial values of the SLSQP algorithm, and are then optimised to minimise the loss function
\begin{equation}\label{eqL}
L=\omega_L\sum_i(U_{\rm EMO}(R_i)-U_{\rm GA}(R_i))^2+\sum_j(E^j_{\rm cal}-E^j_{\rm obs})^2,
\end{equation}
where $i$ indexes the potential points listed in Table \ref{tableGAPEC}, $U_{\rm EMO}$ is the potential energy calculated via EMO function, $U_{\rm GA}$ is the corresponding GA fitted value listed in Table \ref{tableGAPEC}, $j$ indexes the rovibrational levels observed in \cite{LiRbLEVELA1b3C12016}, $E_{\rm cal}$ is the rovibrational energy calculated by solving Eq. \eqref{eqH} \cite{FGH1989} with $\xi_{\rm bb}$ calculated via CI-CPP in \cite{abPECSOCPCCP2020}, $E_{\rm obs}$ is the corresponding experimental value in \cite{LiRbLEVELA1b3C12016}, and $\omega_{L}$ is weight of potential points in the loss function. 

The SLSQP algorithm implimented in SciPy 1.13.0 \cite{optSciPy} is used in the present work. As one of the traditional optimisation algorithms, it tends to be sensitive to the initial value. If the initial value is set inappropriately, it may fall into a local minimum. In comparison, GA is more efficient to search larger parameter space and target the global minimum, but costs more computational resources. Combining these two methods tends to work better, which is the core strategy of our methodology. 

\section{Discussions}
\subsection{\label{Sec3A}Fitted results}
\begin{figure}[htbp]
	\centering
	\includegraphics[width=0.45\textwidth]{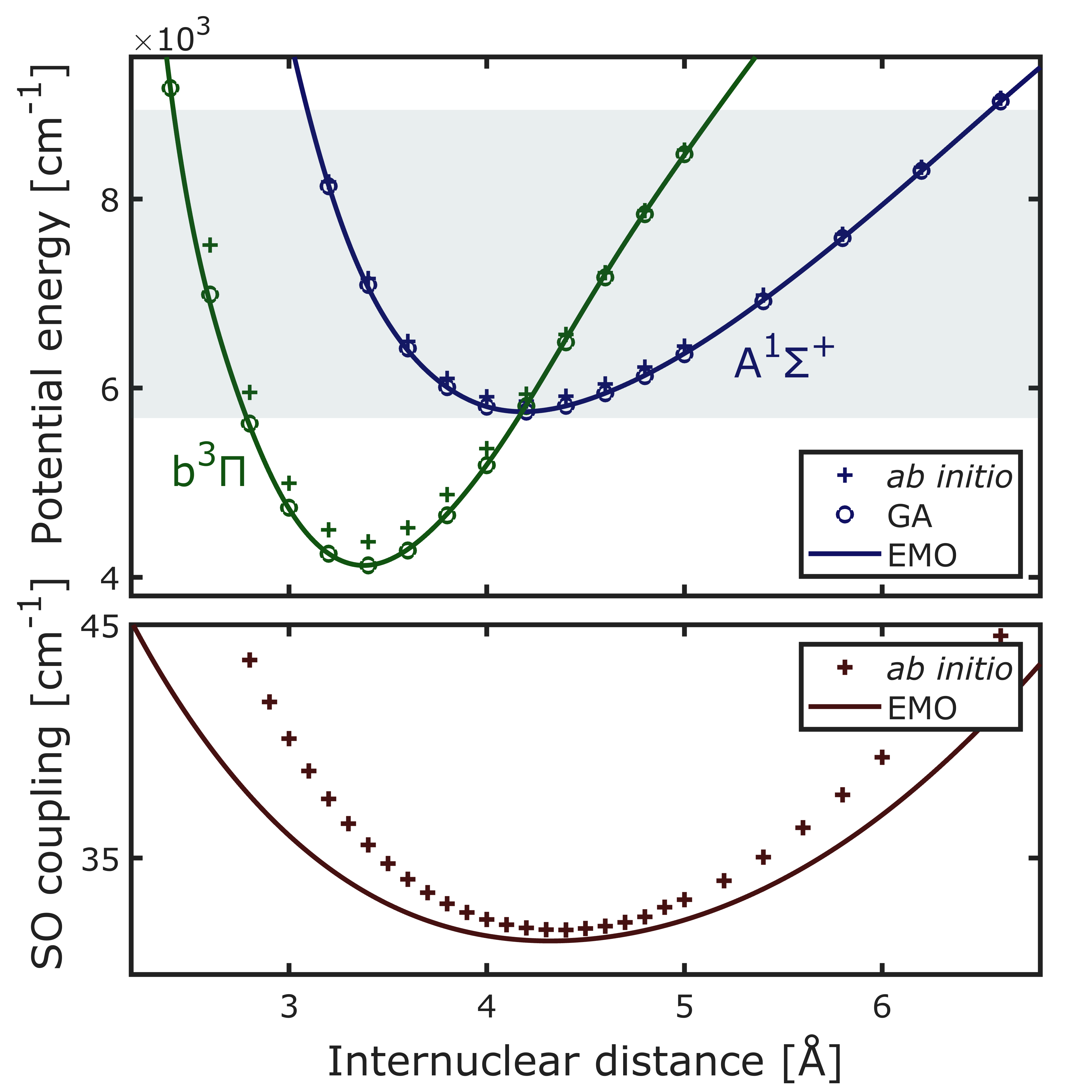}
	\caption{The comparison of $U_{\rm A}$, $U_{\rm b}$ (up panel) and $\xi_{\rm Ab}$ (down panel) obtained by $ab\ initio$ calculations (crosses), GA optimisation (circles) and EMO function fitting (solid lines). The shaded area denotes the energy range of observed rovibrational levels in \cite{LiRbLEVELA1b3C12016}.}
	\label{fitPECsFig}
\end{figure}

We adopt our method on the LiRb molecule. First of all, the $ab\ initio$ point-wise PECs $U_{\rm A}$ and $U_{\rm b}$ are optimised through GA. After repeated trials, the hyperparameters of GA are tuned to $N=1000$, $T=400$, $\omega_{F}=0.002$, $\lambda=0.5$, and $P_{\rm x}=0.3$. The procedure is run 10 times under these hyperparameters. The physically best results are picked from these 10 runs, as listed in Table  \ref{tableGAPEC}. One can see that the point-wise PECs near the bottom of potentials have been optimised about hundreds of $\rm cm^{-1}$. Afterwards, the GA-optimised $U_{\rm A}$ and $U_{\rm b}$ in Table  \ref{tableGAPEC} and $\xi_{\rm Ab}$ calculated in \cite{abPECSOCPCCP2020} are fitted into the EMO function. The coefficient $\omega_L$ in the loss function of SLSQP algorithm is set to 0.0001. The fitted parameters of the EMO functions are listed in Table  \ref{tableEMOPSO}. As shown in Fig. \ref{fitPECsFig}, compared to $ab\ initio$ calculations, the GA-optimised PECs have deeper potential wells, and so do the EMO PECs. The fitted SO coupling curve also has a deeper and wider well compared to $ab\ initio$ calculations. 
\begin{table}[htbp]
	\caption{\label{tableEMOPSO}
		The parameters of EMO PECs and SO coupling for the $\rm A^1\Sigma^+$ and $\rm b^3\Pi$ states of LiRb. The parameters with superscript f are fixed. The remaining parameters are optimised via the SLSQP algorithm. Energies are given in $\rm cm^{-1}$, internuclear distances are in $\rm\AA$ and coefficients $\beta_i$ are in 1/$\rm\AA$.}
	\begin{ruledtabular}
		\begin{tabular}{rrrr}
			& $U_{\rm A}$           & $U_{\rm b}$          & $\xi_{\rm Ab}$   \\\hline
			$q$        & 3\footnotemark[6] & 4\footnotemark[6] & 1\footnotemark[6] \\
			$R_{\rm ref}$  & 4.2\footnotemark[6] & 3.4\footnotemark[6] & 4.6\footnotemark[6] \\\hline
			$T_e$      & 5749.9358       & 4124.2063      & 31.4452      \\
			$D_e$      & 6987.4109       & 8613.1404      & 47.7538      \\
			$R_e$      & 4.1819\footnotemark[6] & 3.3779\footnotemark[6] & 4.3196 \\\hline
			$\beta_0$  & 0.4558156100    & 0.6659361156   & 0.2105957306 \\
			$\beta_1$  & -0.1003385339   & 0.2005301984   & 0.1138233161 \\
			$\beta_2$  & -0.4728226328   & -0.3765692448  & 0.7268579657\footnotemark[6]   \\
			$\beta_3$  & 1.3235443844    & -1.6712335641  & 2.1653083977\footnotemark[6]   \\
			$\beta_4$  & 6.2113954621    & 4.0219039659   & 2.3716270654\footnotemark[6]   \\
			$\beta_5$  & -12.6227039293  & 15.1945920816  &              \\
			$\beta_6$  & -30.6398768819  & -20.5050486971 &              \\
			$\beta_7$  & 60.3403046173   & -54.1544811936 &              \\
			$\beta_8$  & 76.0445287778   & 55.1534757328  &              \\
			$\beta_9$  & -138.5647758852 & 81.8727587255  &              \\
			$\beta_{10}$ & -92.2212822226  & -71.0566484870 &              \\
			$\beta_{11}$ & 152.1359040286  & -45.5314864256 &              \\
			$\beta_{12}$ & 43.8793924613   & 35.0427044818  &              \\
			$\beta_{13}$ & -63.8511533223  & 2.1812779222   &             
		\end{tabular}
	\end{ruledtabular}
\end{table}

\begin{figure}[htbp]
	\centering
	\includegraphics[width=0.45\textwidth]{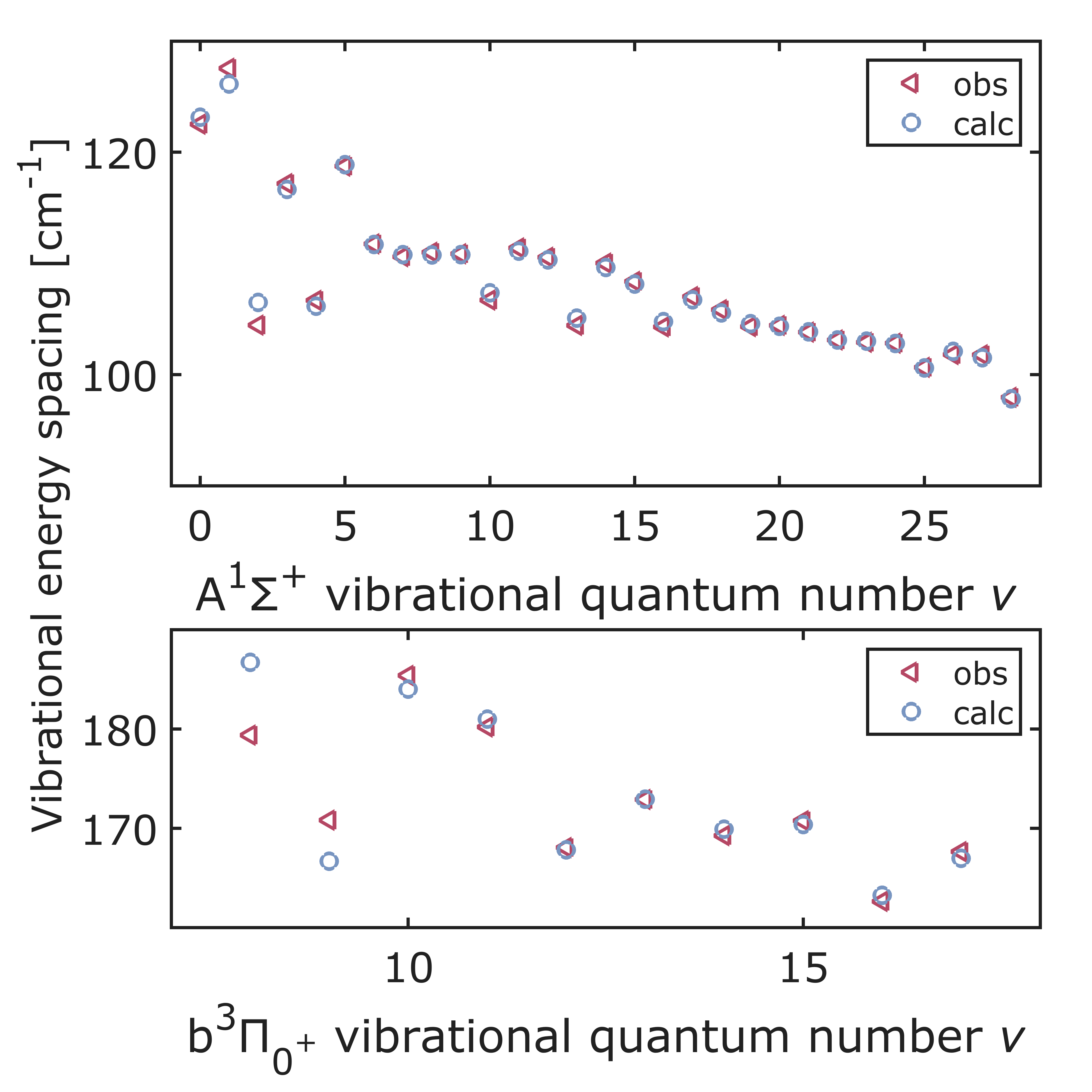}
	\caption{The comparison of experimental vibrational spacings \cite{LiRbLEVELA1b3C12016} (red triangles) to the calculated values with the analytical PECs given in Table \ref{tableEMOPSO} (blue circles) for ${\rm A}^1\Sigma^+$ state (up panel) and ${\rm b}^3\Pi_{0^+}$ state (low panel) of $^7$Li$^{85}$Rb molecule. }
	\label{errorEdFig}
\end{figure}

We calculate the rovibrational energies with the PECs and SO terms obtained by $ab\ initio$ calculations, GA optimisation and EMO function fitting, respectively, as seen in the supplemental materials \footnote[2]{See Supplemental Material at [URL will be inserted by publisher].}. In \cite{LiRbLEVELA1b3C12016}, the energy spacings between the neighbor vibrational levels $\Delta E_v=E_{v+1}-E_{v}$ calculated with the PECs taken from \cite{abPECLB2009} are compared to the experimental observed values. There are nearly a half of energy spacings with an error larger than 1 $\rm cm^{-1}$. Similarly, we also calculate the energy spacings with the EMO PECs and SO coupling, which agree well with values determined by the experimental data, as denoted illustrated in Fig. \ref{errorEdFig}. The root mean squared error between the theoretically calculated energy spacings and their observed values are less than 0.5 $\rm cm^{-1}$, without considering $\rm A^1\Sigma^+$ $v = 2$ level and $\rm b^3\Pi_{0^+}$ $v = 8, 9$ levels. To fit these 3 rovibrational levels more accurately, it may need to consider other components of the $\rm b^3\Pi$ state and the spin-rotational coupling between them.
\subsection{\label{Sec3B}Future prospects}
\begin{figure*}[htbp]
	\centering
	\includegraphics[width=0.9\textwidth]{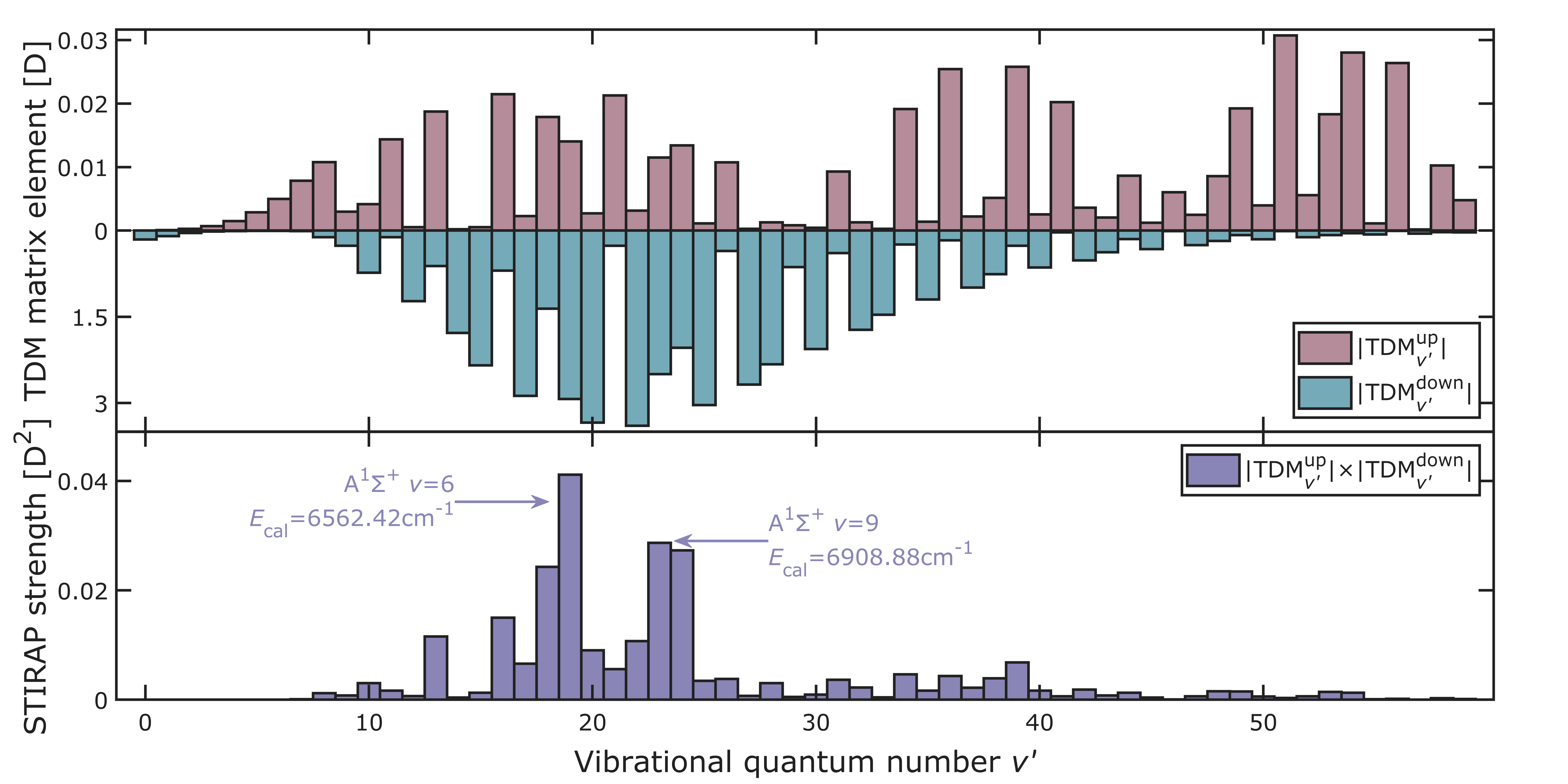}
	\caption{The TDM matrix elements for two legs (up panel) and strengths (down panel) of STIRAP transfer from the Feshbach state to the absolute rovibrational ground state for $^6$Li$^{87}$Rb, with the rovibrational levels of the SO coupled ${\rm A}^1\Sigma^+$ and ${\rm b}^3\Pi_{0^+}$ states as intermediate state. The rotational quantum number $J$ equals 1.}
	\label{TDMsAbFig}
\end{figure*}
The physical model fitted in this work are useful for finding pathway to produce ultracold molecules in the absolute rovibrational ground state. In previous work, Stevenson $et\ al.$ proposed several STIRAP pathways to transfer bosonic $^7$Li$^{85}$Rb molecules to the absolute rovibrational ground state. In \cite{LiRbLEVELA1b3C12016,PALi7Rb852016}, they produced $^7$Li$^{85}$Rb molecules in $\rm X^1\Sigma^+$ $v=43$ level through PA followed by spontaneous decay, and suggested using $\rm C^1\Sigma^+$ $v=22$ level as intermediate state for STIRAP to transfer them to $\rm X^1\Sigma^+$ $v=0$ level. In  \cite{LiRbLEVELd3Pi2016}, they also proposed a STIRAP pathway to transfer Feshbach $^7$Li$^{85}$Rb molecules to the absolute rovibrational ground state, with the SO coupled $\rm D^1\Pi$ and $\rm d^3\Pi$ states as intermediate state. Besides, as illustrated in Fig. \ref{abPECsFig}, the rovibrational levels in the SO coupled $\rm A^1\Sigma^+$ and $\rm b^3\Pi_{0^+}$ states could also be chosen as the intermediate state for STIRAP. One possible STIRAP pathway has been recommended in \cite{ICSThesis2018}, which chooses $\rm A^1\Sigma^+$ $v=5$ level or $\rm b^3\Pi_{0^+}$ $v=12$ level as intermediate state. However, the STIRAP pathway proposed in \cite{ICSThesis2018} is not based on deperturbation analysis, and could not be accurately generalised to other isotopes, such as $^6$Li$^{87}$Rb molecule. With the PECs and SO coupling fitted in this work, we could calculate the TDM matrix elements, which are particularly helpful in predicting feasible intermediate state for STIRAP transfer. 

For fermionic $^{6}$Li$^{87}$Rb, the TDM matrix elements for the transitions from the rovibrational levels of the coupled ${\rm A}^1\Sigma^+$-${\rm b}^3\Pi_{0^+}$ states to the Feshbach state and to the absolute rovibrational ground state are calculated as 
\begin{equation}\label{eqTDM1}
{\rm TDM}_{v'}^{\rm up}=\int\Phi_{\rm F}\cdot{\hat\mu_{\rm ab}}\cdot\Phi_{\rm b0}(v',J=1){\rm d}R,
\end{equation}
and
\begin{equation}\label{eqTDM2}
{\rm TDM}_{v'}^{\rm down}=\int\Phi_{\rm X}\cdot{\hat\mu_{\rm AX}}\cdot\Phi_{\rm A}(v',J=1){\rm d}R,
\end{equation}
where $\Phi_{\rm F}$ is the wave function of the Feshbach state, which is approximated as that for the shallowest bound state of the $\rm a^3\Sigma^+$ state \cite{DPNaRb2007}, $\Phi_{\rm X}$ is the wave function of the absolute rovibrational ground state, $\Phi_{\rm A}$ and $\Phi_{\rm b0}$ are components of the eifenfunction calculated by solving Eq. \eqref{eqH}, ${\hat\mu_{\rm ab}}$ is the $R$-dependent TDM between $\rm a^3\Sigma^+$ state and $\rm b^3\Pi$ state taken from \cite{abTDMPCCP2018}, and ${\hat\mu_{\rm AX}}$ is the $R$-dependent TDM between $\rm X^1\Sigma^+$ state and $\rm A^1\Sigma^+$ state taken from \cite{abTDM2012}. $\Phi_{\rm F}$ and $\Phi_{\rm X}$ are calculated based on the empirical PECs in \cite{TiemannIPAX1Sa3S2011}. $\Phi_{\rm A}$ and $\Phi_{\rm b0}$ are calculated based on the PECs and the SO coupling listed in Table \ref*{tableEMOPSO}, together with $\xi_{\rm bb}$ adopted from \cite{abPECSOCPCCP2020}.
\begin{table}[b]
	\caption{\label{tableTDM}The feasible intermediate states for STIRAP predicted in $\rm A^1\Sigma^+$-$\rm b^3\Pi_{0^+}$ state. The wavelengths of the pump and Stokes laser are denoted as $\lambda_{\rm p}$ and $\lambda_{\rm S}$, respectively. Energies are given in $\rm cm^{-1}$, wavelengths are in nm, and TDM matrix elements are in Debye.}
	\begin{ruledtabular}
	\begin{tabular}{lll}
		& \multicolumn{2}{l}{Feasible intermediate states}    \\\hline
	    & ${\rm A}^1\Sigma^+$ $v=6$ \footnotemark[19]  & ${\rm A}^1\Sigma^+$ $v=9$ \footnotemark[19]  \\
	    & ${\rm b}^3\Pi_{0^+}$ $v=12$ & ${\rm b}^3\Pi_{0^+}$ $v=14$ \\\hline
	$|{\rm TDM}_{v''}^{\rm up}|$ & 0.0140 & 0.0115  \\
	$|{\rm TDM}_{v''}^{\rm down}|$ & 2.932 & 2.498  \\
	$E_{v''}^{\rm CC}$                   & 6562.42                  & 6908.88   \\
	$\lambda_{\rm p}$           & 1523.8                   & 1447.4  \\
	$\lambda_{\rm S}$           & 807.4                    & 785.4                   
	\end{tabular}
	\end{ruledtabular}
	\footnotetext[19]{The rovibrational level is assigned to this electronic state.}
\end{table}

The STIRAP strength is estimated as
\begin{equation}\label{eqI}
I_{v'}=|\rm TDM_{\it v'}^{up}|\cdot|TDM_{\it v'}^{down}|.
\end{equation}
As shown in Fig. \ref{TDMsAbFig}, the $v'=19$ level stands as the best feasible intermediate state for the STIRAP transfer. It is assigned as $\rm A^1\Sigma^+$ $v=6$ level coupled with the $\rm b^3\Pi_{0^+}$ $v=12$ level. In addition, the $v'=23$ level is also recommended as a feasible intermediate state, which is assigned as $\rm A^1\Sigma^+$ $v=9$ level coupled with the $\rm b^3\Pi_{0^+}$ $v=14$ level. Their rovibrational energies and relevant wavelengths of the pump laser and Stokes laser are listed in Table  \ref{tableTDM}. More rovibraional energies calculated for $^{6}$Li$^{87}$Rb are listed in \footnotemark[2].

The SO coupled $\rm B^1\Pi$, $\rm c^3\Sigma^+$ and $\rm b^3\Pi$ states have also been used as intermediate states for the STIRAP transfer \cite{MANa23K402021}. Our methods could be adopted for the deperturbation analysis of the $\rm B^1\Pi$, $\rm c^3\Sigma^+$ and $\rm b^3\Pi$ states. However, there are no relevant spectroscopic data, especially those of $\rm c^3\Sigma^+$ state. Based on the $ab\ initio$ PECs and SO couplings in \cite{abPECSOCPCCP2020}, we also predict the rovibrational levels of the coupled $\rm B^1\Pi$, $\rm c^3\Sigma^+$ and $\rm b^3\Pi$ states that may be suitable as intermediate state for STIRAP in Appendix \ref{App2}. 

\section{Conclusion}
A genetics-based deperturbation analysis on the SO coupled $\rm A^1\Sigma^+$ and $\rm b^3\Pi$ states of LiRb molecule was performed in this work. Utilizing experimentally observed rovibrational levels of the $\rm A^1\Sigma^+$ and $\rm b^3\Pi_{0^+}$ states of $^7$Li$^{85}$Rb, we optimised point-wise PECs of $\rm A^1\Sigma^+$ and $\rm b^3\Pi$ states using GA. To generalise the algorithm in \cite{GALiRb2019} to the CC problem, we introduced the concept of chromosome to encode each PEC. To address the limitation of the observed energy range, we employed a nonlinear fitness function considering both rovibrational levels and potential points. Afterwards, the point-wise potentials were fitted into the EMO functional form. The parameters of EMO functions were further optimised using SLSQP algorithm. 

The fitted underlying interactions are vital for guiding the preparation of ultracold molecules in experiments. Based on the fitted PECs and SO coupling, the TDM matrix elements for transitions from the rovibrational levels of the coupled  $\rm A^1\Sigma^+$ and $\rm b^3\Pi_{0^+}$ states to the Feshbach state and to the absolute rovibrational ground state for fermionic $^6$Li$^{87}$Rb molecule were calculated. We recommended a pathway for STIRAP transfer from the Feshbach state to the absolute rovibrational ground state of $^6$Li$^{87}$Rb with 1523.8 nm pump laser and 807.4 nm Stokes laser. Additionally, we also estimated the feasible rovibrational levels of the SO coupled $\rm B^1\Pi$, $\rm c^3\Sigma^+_1$ and $\rm b^3\Pi_1$ states suitable as intermediate state for STIRAP based on $ab\ initio$ PECs and SO couplings. Although there is an error within 30 $\rm cm^{-1}$, the estimation is sufficient to provide a good starting point for the experimental search for a suitable intermediate state. Moreover, the methodologies developed in this work can be generalized to other electronic states of LiRb and other molecular species.

\begin{acknowledgments}
This work has been supported by National Natural Science Foundation of China (Grants No.12104082 and No.12241409), National Key R\&D Program of China (Grant No.2018YFA0306503), Innovation Program for Quantum Science and Technology (Grant No.2021ZD0302100), and Program of the State Key Laboratory of Quantum Optics and Quantum Optics Devices (No.KF201814). Xin-Yu Luo is funded by the European Union (ERC, DiMoBecTe, 101125173). Xin-Yu Luo gratefully acknowledge support from the Max Planck Society, and the Deutsche Forschungsgemeinschaft under Germany's Excellence Strategy -- EXC-2111 -- 390814868. 
\end{acknowledgments}

\appendix

\section{The fitness terms concerning potential points and rovibrational levels}\label{App1}
To define $F_\alpha$ concerning potential points, several potential points in the short-range and long-range regions are taken into consideration, as illustrated by the green points in Fig. \ref{iniPop}. These points are out of the experimentally observed energy range, and are defined as reference points. For each individual, $F_\alpha$ is defined as
\begin{equation}\label{eqA3}
	F_\alpha=\sqrt{\sum_s\sum_i\zeta^s_i\left(U^s_{\rm GA}(R_i)-U^s_{\rm ref}(R_i)\right)^2},
\end{equation}
where $i$ indexes reference points, $s$ indexes electronic states, $U^s_{\rm GA}$ is the potential energy encoded in the considered individual, $U_{\rm ref}^s$ is its corresponding theoretical value calculated via CI-CPP in \cite{abPECSOCPCCP2020}, and $\zeta^s_i$ is the weight of different reference points. $\zeta^s_i$ is defined as
\begin{equation}\label{eqA2} 
	\zeta^s_i=\frac{\frac{1}{\sigma^{s}(R_i)}}{\sum_s\sum_i\frac{1}{\sigma^{s}(R_i)}},
\end{equation}
where $\sigma^{s}(R_i)$ is the standard deviation of the potential energy at reference point $R_i$ in the initial population. $\sigma^{s}(R_i)$ is evaluated as
\begin{equation}\label{eqA1}
\sigma^{s}(R_i)=\sqrt{\frac{1}{N}\sum_{n=1}^N(U^s_n(R_i)-{\bar U^s}(R_i))^2},
\end{equation}
where $n$ indexes individuals, $U_n^s(R_i)$ is the potential energy of $s$ electronic state at reference point $R_i$ of the $n$-th individual, and ${\bar U}^s(R_i)$ is the average among all individuals in the population for the potential energy at reference point $R_i$ of $s$ electronic state.

To define $F_\beta$ concerning the rovibraional levels, the positions of some additional rovibrational levels at the middle and bottom of potentials are predicted according to Eq. \eqref{eqTv}, including $\rm A^1\Sigma^+$ $v=33, 34, 36, 37$ levels and $\rm b^3\Pi_{0^+}$ $v=0\text{-}4, 19, 21\text{-}23, 25, 27, 29, 31, 33$ levels. Together with the rovibrational levels observed in \cite{LiRbLEVELA1b3C12016}, there are a total of 59 reference levels taken in account in the fitness function. For each individual, $F_\beta$ is defined as 
\begin{equation}\label{eqA6}
F_\beta=\frac{1}{2}\sum_s\sqrt{\frac{1}{N_{\rm L}^s}{\sum_{j=1}^{N_{\rm L}^s}\left(E^{sj}_{\rm GA}-E^{sj}_{\rm ref}\right)^2}},
\end{equation}
where $j$ indexes reference levels, $N_{\rm L}^s$ is the number of reference levels belonging to the $s$ state, $E^{sj}_{\rm GA}$ is the rovibrational energy calculated by solving Eq. \eqref{eqH} with PECs encoded in current individual and $\xi_{\rm Ab}$ and $\xi_{\rm bb}$ calculated via CI-CPP in \cite{abPECSOCPCCP2020}, and $E^{sj}_{\rm ref}$ is the reference value observed experimentally in \cite{LiRbLEVELA1b3C12016} or predicted via Eq. \eqref{eqTv}. 

\section{The estimation of $\rm B^1\Pi$, $\rm c^3\Sigma^+$ and $\rm b^3\Pi$ states as the intermediate state for STIRAP}\label{App2}
\begin{figure*}[htbp]
	\centering
	\includegraphics[width=0.9\textwidth]{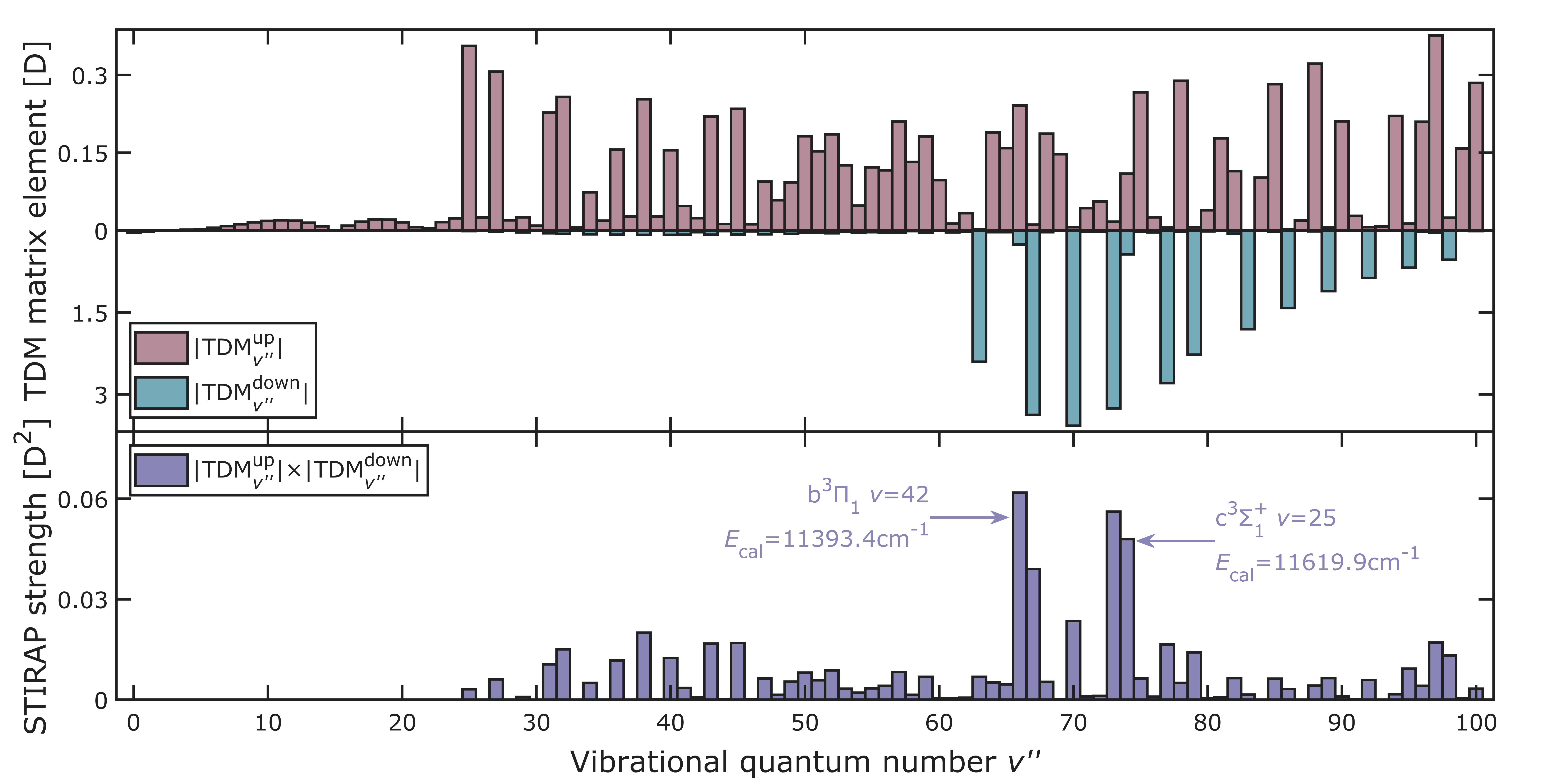}
	\caption{The TDM matrix elements for two legs (up panel) and strengths (down panel) for STIRAP transfer from the Feshbach state to the absolute rovibrational ground state for $^6$Li$^{87}$Rb, with the rovibrational levels of the SO coupled ${\rm B}^1\Pi$, ${\rm b}^3\Pi$ and ${\rm c}^3\Sigma^+$ states as intermediate state. The rotational quantum number $J$ equals to 1.}
	\label{TDMsBcbFig}
\end{figure*}
For the SO coupled ${\rm B}^1\Pi$, ${\rm c}^3\Sigma^+$ and ${\rm b}^3\Pi$ states, one should typically consider their components of $\Omega=1$, which are strongly coupled via SO coupling \cite{MANaCsNi2023}. The relevant CC Schrödinger equation is
\begin{widetext}
\begin{equation}
\begin{array}{ll}
	\left(\begin{array}{lll}
		T+U_{\rm B}+E^{\rm R}_{\rm B} & \xi_{\rm Bc} & -\xi_{\rm Bb}\\
		\xi_{\rm Bc}  & T+U_{\rm c}+E^{\rm R}_{\rm c1} & \xi_{\rm bc} \\
		-\xi_{\rm Bb}  &\xi_{\rm bc}   & T+U_{\rm b}+E^{\rm R}_{\rm b1}\\
	\end{array}\right)
	\left(\begin{array}{lll}
		\Phi_{\rm B} \\\Phi_{\rm c1} \\\Phi_{\rm b1}\\
	\end{array}\right)
	=E^{\rm CC}_{v''}
	\left(\begin{array}{ll}
		\Phi_{\rm B} \\\Phi_{\rm c1} \\\Phi_{\rm b1}\\
	\end{array}\right),\\
\end{array}
\end{equation}
\end{widetext}
where $U_{\rm B}$, $U_{\rm c}$ and $U_{\rm b}$ stand for the diabatic PECs of the $\rm B^1\Pi$, $\rm c^3\Sigma^+$ and $\rm b^3\Pi$ states, respectively, $\xi_{\rm Bc}$, $\xi_{\rm Bb}$ and $\xi_{\rm bc}$ are the SO coupling terms, $v''$ indexes the rovibrational levels of the coupled state, $E_{v''}^{\rm CC}$ is the rovibrational energy, and $\Phi_{\rm B}$, $\Phi_{\rm c1}$ and $\Phi_{\rm b1}$ are the components of the eigenstate belonging to the $\rm B^1\Pi$, $\rm c^3\Sigma^+_1$ and $\rm b^3\Pi_1$ states, respectively. According to Eq. \eqref{eqROT1}, the rotational Hamiltonians are
\begin{equation}
\begin{array}{lll}
	E^{\rm R}_{\rm b1}=	E^{\rm R}_{\rm c1}=\frac{\hbar^2}{2\mu R^2}\left[J(J+1)+2\right],\\
	E^{\rm R}_{\rm B}=\frac{\hbar^2}{2\mu R^2}\left[J(J+1)\right].\\
\end{array}
\end{equation}
Similar to the treatment in Section II A, the eigenstate is assigned to ${\rm B}^1\Pi$, $\rm c^3\Sigma^+_1$ or ${\rm b}^3\Pi_1$ state by comparing their fractional partitions. 

Since the lack of spectroscopic data of $\rm c^3\Sigma^+_1$ and ${\rm b}^3\Pi_1$ states, we cannot perform the deperturbation analysis. Instead, we use PECs and SO couplings calculated by CI-CPP in \cite{abPECSOCPCCP2020} to predict the feasible rovibational levels of the coupled $\rm B^1\Pi$, $\rm c^3\Sigma^+_1$ and $\rm b^3\Pi_1$ states suitable as the intermediate state for STIRAP. The error of used PECs could be roughly estimated by comparing with the data of ${\rm B}^1\Pi$ rovibrational levels observed in \cite{TiemannLEVELB1PiD1Pi2013}. Based on the PECs and SO coupling in \cite{abPECSOCPCCP2020}, the rovibrational energy of ${\rm B}^1\Pi$ $v=3$, $J=1$ level for $^7$Li$^{85}$Rb is calculated as 11594.3 $\rm cm^{-1}$, and that of ${\rm B}^1\Pi$ $v=11$, $J=1$ level for $^7$Li$^{87}$Rb is calculated as 12190.9 $\rm cm^{-1}$. According to the spectroscopic data in \cite{TiemannLEVELB1PiD1Pi2013}, their experimentally observed values are 11568.5 $\rm cm^{-1}$ and 12170.6 $\rm cm^{-1}$, respectively. Therefore, the rovibrational energies calculated based on $ab\ initio$ PECs are higher than the real value, and the errors are within 30 $\rm cm^{-1}$.
\begin{table}[htbp]
	\caption{\label{tableFCF}The possible intermediate states for STIRAP predicted in $\rm B^1\Pi$-$\rm c^3\Sigma^+_1$-$\rm b^3\Pi_1$ state. The wavelengths of the pump and Stokes laser are denoted as $\lambda_{\rm p}$ and $\lambda_{\rm S}$, respectively. Energies are given in $\rm cm^{-1}$, wavelengths are in nm, and TDM matrix elements are in Debye.}
	\begin{ruledtabular}
		\begin{tabular}{lll}
			& \multicolumn{2}{l}{Possible intermediate states}    \\\hline
			& ${\rm b}^3\Pi_1$ $v=42$ \footnotemark[19]   & ${\rm c}^3\Sigma^+_1$ $v=25$ \footnotemark[19]      \\
			& ${\rm B}^1\Pi$ $v=1$ & ${\rm B}^1\Pi$ $v=1$     \\\hline
			$|{\rm TDM}_{v''}^{\rm up}|$   & 0.2414              &  0.1099            \\
			$|{\rm TDM}_{v''}^{\rm down}|$ & 0.2562                & 0.4365            \\
			$E_{v''}^{\rm CC}$     & 11393.4                  & 11619.9                  \\
			$\lambda_p$            & 860.6                    & 877.7                    \\
			$\lambda_S$            & 573.3                    & 580.8                   
		\end{tabular}
	\end{ruledtabular}
	\footnotetext[19]{The rovibrational level is assigned to this electronic state.}
\end{table}

We calculate TDM matrix elements for two legs of the STIRAP transfer, as follows: 
\begin{equation}
	\begin{array}{rr}
	{\rm TDM}^{\rm up}_{v''} =\int \Phi_{\rm F}\cdot{\hat\mu_{\rm ab}}\cdot\Phi_{\rm b1}(v'',J=1){\rm d}R\\+\int \Phi_{\rm F}\cdot{\hat\mu_{\rm ac}}\cdot\Phi_{\rm c1}(v'',J=1){\rm d}R,
	\end{array}
\end{equation}
\begin{equation}
{\rm TDM}^{\rm down}_{v''} = \int \Phi_{\rm X}\cdot{\hat\mu_{\rm BX}}\cdot\Phi_{\rm B}(v'',J=1){\rm d}R,
\end{equation}
where ${\hat\mu_{\rm BX}}$ and ${\hat\mu_{\rm ac}}$ are $R$-dependent TDMs from $\rm X^1\Sigma^+$ state to $\rm B^1\Pi$ state and from $\rm a^3\Sigma^+$ state to $\rm c^3\Sigma^+$ state, respectively, both of which are taken from \cite{abTDMPCCP2018}. The STIRAP strength is estimated as 
\begin{equation}
I_{v''}=\left|{\rm TDM}^{\rm up}_{v''}\right|\cdot\left|{\rm TDM}^{\rm down}_{v''}\right|.
\end{equation}
The TDM matrix elements and STIRAP strengths are shown in Fig. \ref{TDMsBcbFig}. As seen, one possible intermediate state for STIRAP transfer is the ${\rm c}^3\Sigma_1^+$ $v=25$ level, which is coupled to ${\rm B}^1\Pi$ $v=3$ level. Another possible intermediate state for STIRAP transfer is the ${\rm b}^3\Pi_1$ $v=42$ level coupled to ${\rm B}^1\Pi$ $v=1$ level. Their rovibraional energies and relevant wavelengths of pump laser and Stocks laser are listed in Table \ref{tableFCF}. Take the ${\rm c}^3\Sigma_1^+$ $v=25$ level coupled to ${\rm B}^1\Pi$ $v=3$ level as an example, considering there is an error of 30 $\rm cm^{-1}$ higher than the real value, the rovibrational energy should be 11589.9 $\rm cm^{-1}$. The corresponding wavelengths of pump laser and Stocks laser are 862.8 nm and 574.3 nm, respectively. The error here is less than 3 nm, which is easily covered by tunable wavelength ranges of semiconductor or Ti:Sappire lasers. 

\nocite{*}

\bibliographystyle{apsrev4-2}
\bibliography{Refs} 

\begin{thebibliography}{81}%
\makeatletter
\providecommand \@ifxundefined [1]{%
 \@ifx{#1\undefined}
}%
\providecommand \@ifnum [1]{%
 \ifnum #1\expandafter \@firstoftwo
 \else \expandafter \@secondoftwo
 \fi
}%
\providecommand \@ifx [1]{%
 \ifx #1\expandafter \@firstoftwo
 \else \expandafter \@secondoftwo
 \fi
}%
\providecommand \natexlab [1]{#1}%
\providecommand \enquote  [1]{``#1''}%
\providecommand \bibnamefont  [1]{#1}%
\providecommand \bibfnamefont [1]{#1}%
\providecommand \citenamefont [1]{#1}%
\providecommand \href@noop [0]{\@secondoftwo}%
\providecommand \href [0]{\begingroup \@sanitize@url \@href}%
\providecommand \@href[1]{\@@startlink{#1}\@@href}%
\providecommand \@@href[1]{\endgroup#1\@@endlink}%
\providecommand \@sanitize@url [0]{\catcode `\\12\catcode `\$12\catcode
  `\&12\catcode `\#12\catcode `\^12\catcode `\_12\catcode `\%12\relax}%
\providecommand \@@startlink[1]{}%
\providecommand \@@endlink[0]{}%
\providecommand \url  [0]{\begingroup\@sanitize@url \@url }%
\providecommand \@url [1]{\endgroup\@href {#1}{\urlprefix }}%
\providecommand \urlprefix  [0]{URL }%
\providecommand \Eprint [0]{\href }%
\providecommand \doibase [0]{https://doi.org/}%
\providecommand \selectlanguage [0]{\@gobble}%
\providecommand \bibinfo  [0]{\@secondoftwo}%
\providecommand \bibfield  [0]{\@secondoftwo}%
\providecommand \translation [1]{[#1]}%
\providecommand \BibitemOpen [0]{}%
\providecommand \bibitemStop [0]{}%
\providecommand \bibitemNoStop [0]{.\EOS\space}%
\providecommand \EOS [0]{\spacefactor3000\relax}%
\providecommand \BibitemShut  [1]{\csname bibitem#1\endcsname}%
\let\auto@bib@innerbib\@empty
\bibitem [{\citenamefont {Zhai}(2021)}]{BookHuiZhai2021}%
  \BibitemOpen
  \bibfield  {author} {\bibinfo {author} {\bibfnamefont {H.}~\bibnamefont
  {Zhai}},\ }\href {https://doi.org/10.1017/9781108595216} {\emph {\bibinfo
  {title} {Ultracold Atomic Physics}}}\ (\bibinfo  {publisher} {Cambridge
  University Press},\ \bibinfo {year} {2021})\BibitemShut {NoStop}%
\bibitem [{\citenamefont {Pérez-Ríos}(2020)}]{BookJesusRios2020}%
  \BibitemOpen
  \bibfield  {author} {\bibinfo {author} {\bibfnamefont {J.}~\bibnamefont
  {Pérez-Ríos}},\ }\href {https://doi.org/10.1007/978-3-030-55936-6} {\emph
  {\bibinfo {title} {An Introduction to Cold and Ultracold Chemistry}}}\
  (\bibinfo  {publisher} {Springer International Publishing, New York},\
  \bibinfo {year} {2020})\BibitemShut {NoStop}%
\bibitem [{\citenamefont {Jones}\ \emph {et~al.}(2006)\citenamefont {Jones},
  \citenamefont {Tiesinga}, \citenamefont {Lett},\ and\ \citenamefont
  {Julienne}}]{Rev1PA2006}%
  \BibitemOpen
  \bibfield  {author} {\bibinfo {author} {\bibfnamefont {K.~M.}\ \bibnamefont
  {Jones}}, \bibinfo {author} {\bibfnamefont {E.}~\bibnamefont {Tiesinga}},
  \bibinfo {author} {\bibfnamefont {P.~D.}\ \bibnamefont {Lett}},\ and\
  \bibinfo {author} {\bibfnamefont {P.~S.}\ \bibnamefont {Julienne}},\ }\href
  {https://doi.org/10.1103/RevModPhys.78.483} {\bibfield  {journal} {\bibinfo
  {journal} {REVIEWS OF MODERN PHYSICS}\ }\textbf {\bibinfo {volume} {78}},\
  \bibinfo {pages} {483} (\bibinfo {year} {2006})}\BibitemShut {NoStop}%
\bibitem [{\citenamefont {Ulmanis}\ \emph {et~al.}(2012)\citenamefont
  {Ulmanis}, \citenamefont {Deiglmayr}, \citenamefont {Repp}, \citenamefont
  {Wester},\ and\ \citenamefont {Weidemueller}}]{Rev2PA12012}%
  \BibitemOpen
  \bibfield  {author} {\bibinfo {author} {\bibfnamefont {J.}~\bibnamefont
  {Ulmanis}}, \bibinfo {author} {\bibfnamefont {J.}~\bibnamefont {Deiglmayr}},
  \bibinfo {author} {\bibfnamefont {M.}~\bibnamefont {Repp}}, \bibinfo {author}
  {\bibfnamefont {R.}~\bibnamefont {Wester}},\ and\ \bibinfo {author}
  {\bibfnamefont {M.}~\bibnamefont {Weidemueller}},\ }\href
  {https://doi.org/10.1021/cr300215h} {\bibfield  {journal} {\bibinfo
  {journal} {CHEMICAL REVIEWS}\ }\textbf {\bibinfo {volume} {112}},\ \bibinfo
  {pages} {4890} (\bibinfo {year} {2012})}\BibitemShut {NoStop}%
\bibitem [{\citenamefont {Krems}\ \emph {et~al.}(2009)\citenamefont {Krems},
  \citenamefont {Friedrich},\ and\ \citenamefont {Stwalley}}]{BookColdMol2009}%
  \BibitemOpen
  \bibfield  {author} {\bibinfo {author} {\bibfnamefont {R.}~\bibnamefont
  {Krems}}, \bibinfo {author} {\bibfnamefont {B.}~\bibnamefont {Friedrich}},\
  and\ \bibinfo {author} {\bibfnamefont {W.~C.}\ \bibnamefont {Stwalley}},\
  }\href@noop {} {\emph {\bibinfo {title} {Cold Molecules: Theory, Experiment,
  Applications}}}\ (\bibinfo  {publisher} {CRC press},\ \bibinfo {year}
  {2009})\BibitemShut {NoStop}%
\bibitem [{\citenamefont {Chin}\ \emph {et~al.}(2010)\citenamefont {Chin},
  \citenamefont {Grimm}, \citenamefont {Julienne},\ and\ \citenamefont
  {Tiesinga}}]{RevFB2010}%
  \BibitemOpen
  \bibfield  {author} {\bibinfo {author} {\bibfnamefont {C.}~\bibnamefont
  {Chin}}, \bibinfo {author} {\bibfnamefont {R.}~\bibnamefont {Grimm}},
  \bibinfo {author} {\bibfnamefont {P.}~\bibnamefont {Julienne}},\ and\
  \bibinfo {author} {\bibfnamefont {E.}~\bibnamefont {Tiesinga}},\ }\href
  {https://doi.org/10.1103/RevModPhys.82.1225} {\bibfield  {journal} {\bibinfo
  {journal} {REVIEWS OF MODERN PHYSICS}\ }\textbf {\bibinfo {volume} {82}},\
  \bibinfo {pages} {1225} (\bibinfo {year} {2010})}\BibitemShut {NoStop}%
\bibitem [{\citenamefont {Vitanov}\ \emph {et~al.}(2017)\citenamefont
  {Vitanov}, \citenamefont {Rangelov}, \citenamefont {Shore},\ and\
  \citenamefont {Bergmann}}]{RevSTIRAP2017}%
  \BibitemOpen
  \bibfield  {author} {\bibinfo {author} {\bibfnamefont {N.~V.}\ \bibnamefont
  {Vitanov}}, \bibinfo {author} {\bibfnamefont {A.~A.}\ \bibnamefont
  {Rangelov}}, \bibinfo {author} {\bibfnamefont {B.~W.}\ \bibnamefont
  {Shore}},\ and\ \bibinfo {author} {\bibfnamefont {K.}~\bibnamefont
  {Bergmann}},\ }\href {https://doi.org/10.1103/RevModPhys.89.015006}
  {\bibfield  {journal} {\bibinfo  {journal} {REVIEWS OF MODERN PHYSICS}\
  }\textbf {\bibinfo {volume} {89}},\ \bibinfo {pages} {015006} (\bibinfo
  {year} {2017})}\BibitemShut {NoStop}%
\bibitem [{\citenamefont {Rvachov}\ \emph {et~al.}(2017)\citenamefont
  {Rvachov}, \citenamefont {Son}, \citenamefont {Sommer}, \citenamefont
  {Ebadi}, \citenamefont {Park}, \citenamefont {Zwierlein}, \citenamefont
  {Ketterle},\ and\ \citenamefont {Jamison}}]{MA6Li23Na2017}%
  \BibitemOpen
  \bibfield  {author} {\bibinfo {author} {\bibfnamefont {T.~M.}\ \bibnamefont
  {Rvachov}}, \bibinfo {author} {\bibfnamefont {H.}~\bibnamefont {Son}},
  \bibinfo {author} {\bibfnamefont {A.~T.}\ \bibnamefont {Sommer}}, \bibinfo
  {author} {\bibfnamefont {S.}~\bibnamefont {Ebadi}}, \bibinfo {author}
  {\bibfnamefont {J.~J.}\ \bibnamefont {Park}}, \bibinfo {author}
  {\bibfnamefont {M.~W.}\ \bibnamefont {Zwierlein}}, \bibinfo {author}
  {\bibfnamefont {W.}~\bibnamefont {Ketterle}},\ and\ \bibinfo {author}
  {\bibfnamefont {A.~O.}\ \bibnamefont {Jamison}},\ }\href
  {https://doi.org/10.1103/PhysRevLett.119.143001} {\bibfield  {journal}
  {\bibinfo  {journal} {Phys. Rev. Lett.}\ }\textbf {\bibinfo {volume} {119}},\
  \bibinfo {pages} {143001} (\bibinfo {year} {2017})}\BibitemShut {NoStop}%
\bibitem [{\citenamefont {He}\ \emph {et~al.}(2024)\citenamefont {He},
  \citenamefont {Nie}, \citenamefont {Avalos}, \citenamefont {Botsi},
  \citenamefont {Kumar}, \citenamefont {Yang},\ and\ \citenamefont
  {Dieckmann}}]{MALiK2024reprint}%
  \BibitemOpen
  \bibfield  {author} {\bibinfo {author} {\bibfnamefont {C.}~\bibnamefont
  {He}}, \bibinfo {author} {\bibfnamefont {X.}~\bibnamefont {Nie}}, \bibinfo
  {author} {\bibfnamefont {V.}~\bibnamefont {Avalos}}, \bibinfo {author}
  {\bibfnamefont {S.}~\bibnamefont {Botsi}}, \bibinfo {author} {\bibfnamefont
  {S.}~\bibnamefont {Kumar}}, \bibinfo {author} {\bibfnamefont
  {A.}~\bibnamefont {Yang}},\ and\ \bibinfo {author} {\bibfnamefont
  {K.}~\bibnamefont {Dieckmann}},\ }\href@noop {} {\bibfield  {journal}
  {\bibinfo  {journal} {arXiv:2310.03300v1}\ } (\bibinfo {year}
  {2024})}\BibitemShut {NoStop}%
\bibitem [{\citenamefont {Voges}\ \emph {et~al.}(2020)\citenamefont {Voges},
  \citenamefont {Gersema}, \citenamefont {zum Alten~Borgloh}, \citenamefont
  {Schulze}, \citenamefont {Hartmann}, \citenamefont {Zenesini},\ and\
  \citenamefont {Ospelkaus}}]{MANa23K392020}%
  \BibitemOpen
  \bibfield  {author} {\bibinfo {author} {\bibfnamefont {K.~K.}\ \bibnamefont
  {Voges}}, \bibinfo {author} {\bibfnamefont {P.}~\bibnamefont {Gersema}},
  \bibinfo {author} {\bibfnamefont {M.~M.}\ \bibnamefont {zum Alten~Borgloh}},
  \bibinfo {author} {\bibfnamefont {T.~A.}\ \bibnamefont {Schulze}}, \bibinfo
  {author} {\bibfnamefont {T.}~\bibnamefont {Hartmann}}, \bibinfo {author}
  {\bibfnamefont {A.}~\bibnamefont {Zenesini}},\ and\ \bibinfo {author}
  {\bibfnamefont {S.}~\bibnamefont {Ospelkaus}},\ }\href
  {https://doi.org/10.1103/PhysRevLett.125.083401} {\bibfield  {journal}
  {\bibinfo  {journal} {PHYSICAL REVIEW LETTERS}\ }\textbf {\bibinfo {volume}
  {125}},\ \bibinfo {pages} {083401} (\bibinfo {year} {2020})}\BibitemShut
  {NoStop}%
\bibitem [{\citenamefont {Park}\ \emph {et~al.}(2015)\citenamefont {Park},
  \citenamefont {Will},\ and\ \citenamefont {Zwierlein}}]{MANa23K402015}%
  \BibitemOpen
  \bibfield  {author} {\bibinfo {author} {\bibfnamefont {J.~W.}\ \bibnamefont
  {Park}}, \bibinfo {author} {\bibfnamefont {S.~A.}\ \bibnamefont {Will}},\
  and\ \bibinfo {author} {\bibfnamefont {M.~W.}\ \bibnamefont {Zwierlein}},\
  }\href {https://doi.org/10.1103/PhysRevLett.114.205302} {\bibfield  {journal}
  {\bibinfo  {journal} {Phys. Rev. Lett.}\ }\textbf {\bibinfo {volume} {114}},\
  \bibinfo {pages} {205302} (\bibinfo {year} {2015})}\BibitemShut {NoStop}%
\bibitem [{\citenamefont {See\ss{}elberg}\ \emph {et~al.}(2018)\citenamefont
  {See\ss{}elberg}, \citenamefont {Buchheim}, \citenamefont {Lu}, \citenamefont
  {Schneider}, \citenamefont {Luo}, \citenamefont {Tiemann}, \citenamefont
  {Bloch},\ and\ \citenamefont {Gohle}}]{MANa23K402018}%
  \BibitemOpen
  \bibfield  {author} {\bibinfo {author} {\bibfnamefont {F.}~\bibnamefont
  {See\ss{}elberg}}, \bibinfo {author} {\bibfnamefont {N.}~\bibnamefont
  {Buchheim}}, \bibinfo {author} {\bibfnamefont {Z.-K.}\ \bibnamefont {Lu}},
  \bibinfo {author} {\bibfnamefont {T.}~\bibnamefont {Schneider}}, \bibinfo
  {author} {\bibfnamefont {X.-Y.}\ \bibnamefont {Luo}}, \bibinfo {author}
  {\bibfnamefont {E.}~\bibnamefont {Tiemann}}, \bibinfo {author} {\bibfnamefont
  {I.}~\bibnamefont {Bloch}},\ and\ \bibinfo {author} {\bibfnamefont
  {C.}~\bibnamefont {Gohle}},\ }\href
  {https://doi.org/10.1103/PhysRevA.97.013405} {\bibfield  {journal} {\bibinfo
  {journal} {Phys. Rev. A}\ }\textbf {\bibinfo {volume} {97}},\ \bibinfo
  {pages} {013405} (\bibinfo {year} {2018})}\BibitemShut {NoStop}%
\bibitem [{\citenamefont {Liu}\ \emph {et~al.}(2019)\citenamefont {Liu},
  \citenamefont {Zhang}, \citenamefont {Yang}, \citenamefont {Liu},
  \citenamefont {Nan}, \citenamefont {Rui}, \citenamefont {Zhao},\ and\
  \citenamefont {Pan}}]{MANa23K402019}%
  \BibitemOpen
  \bibfield  {author} {\bibinfo {author} {\bibfnamefont {L.}~\bibnamefont
  {Liu}}, \bibinfo {author} {\bibfnamefont {D.-C.}\ \bibnamefont {Zhang}},
  \bibinfo {author} {\bibfnamefont {H.}~\bibnamefont {Yang}}, \bibinfo {author}
  {\bibfnamefont {Y.-X.}\ \bibnamefont {Liu}}, \bibinfo {author} {\bibfnamefont
  {J.}~\bibnamefont {Nan}}, \bibinfo {author} {\bibfnamefont {J.}~\bibnamefont
  {Rui}}, \bibinfo {author} {\bibfnamefont {B.}~\bibnamefont {Zhao}},\ and\
  \bibinfo {author} {\bibfnamefont {J.-W.}\ \bibnamefont {Pan}},\ }\href
  {https://doi.org/10.1103/PhysRevLett.122.253201} {\bibfield  {journal}
  {\bibinfo  {journal} {Phys. Rev. Lett.}\ }\textbf {\bibinfo {volume} {122}},\
  \bibinfo {pages} {253201} (\bibinfo {year} {2019})}\BibitemShut {NoStop}%
\bibitem [{\citenamefont {Bause}\ \emph {et~al.}(2021)\citenamefont {Bause},
  \citenamefont {Kamijo}, \citenamefont {Chen}, \citenamefont {Duda},
  \citenamefont {Schindewolf}, \citenamefont {Bloch},\ and\ \citenamefont
  {Luo}}]{MANa23K402021}%
  \BibitemOpen
  \bibfield  {author} {\bibinfo {author} {\bibfnamefont {R.}~\bibnamefont
  {Bause}}, \bibinfo {author} {\bibfnamefont {A.}~\bibnamefont {Kamijo}},
  \bibinfo {author} {\bibfnamefont {X.-Y.}\ \bibnamefont {Chen}}, \bibinfo
  {author} {\bibfnamefont {M.}~\bibnamefont {Duda}}, \bibinfo {author}
  {\bibfnamefont {A.}~\bibnamefont {Schindewolf}}, \bibinfo {author}
  {\bibfnamefont {I.}~\bibnamefont {Bloch}},\ and\ \bibinfo {author}
  {\bibfnamefont {X.-Y.}\ \bibnamefont {Luo}},\ }\href
  {https://doi.org/10.1103/PhysRevA.104.043321} {\bibfield  {journal} {\bibinfo
   {journal} {PHYSICAL REVIEW A}\ }\textbf {\bibinfo {volume} {104}},\ \bibinfo
  {pages} {043321} (\bibinfo {year} {2021})}\BibitemShut {NoStop}%
\bibitem [{\citenamefont {Li}\ \emph {et~al.}(2023)\citenamefont {Li},
  \citenamefont {Gu}, \citenamefont {Wang},\ and\ \citenamefont
  {Zhang}}]{MANa23K402023}%
  \BibitemOpen
  \bibfield  {author} {\bibinfo {author} {\bibfnamefont {Z.-L.}\ \bibnamefont
  {Li}}, \bibinfo {author} {\bibfnamefont {Z.-Y.}\ \bibnamefont {Gu}}, \bibinfo
  {author} {\bibfnamefont {P.-J.}\ \bibnamefont {Wang}},\ and\ \bibinfo
  {author} {\bibfnamefont {J.}~\bibnamefont {Zhang}},\ }\href
  {https://doi.org/10.1007/s11433-023-2148-8} {\bibfield  {journal} {\bibinfo
  {journal} {Science China Physics, Mechanics \& Astronomy}\ }\textbf {\bibinfo
  {volume} {66}},\ \bibinfo {pages} {293011} (\bibinfo {year}
  {2023})}\BibitemShut {NoStop}%
\bibitem [{\citenamefont {Guo}\ \emph {et~al.}(2016)\citenamefont {Guo},
  \citenamefont {Zhu}, \citenamefont {Lu}, \citenamefont {Ye}, \citenamefont
  {Wang}, \citenamefont {Vexiau}, \citenamefont {Bouloufa-Maafa}, \citenamefont
  {Quemener}, \citenamefont {Dulieu},\ and\ \citenamefont
  {Wang}}]{MA1Na23Rb872016}%
  \BibitemOpen
  \bibfield  {author} {\bibinfo {author} {\bibfnamefont {M.}~\bibnamefont
  {Guo}}, \bibinfo {author} {\bibfnamefont {B.}~\bibnamefont {Zhu}}, \bibinfo
  {author} {\bibfnamefont {B.}~\bibnamefont {Lu}}, \bibinfo {author}
  {\bibfnamefont {X.}~\bibnamefont {Ye}}, \bibinfo {author} {\bibfnamefont
  {F.}~\bibnamefont {Wang}}, \bibinfo {author} {\bibfnamefont {R.}~\bibnamefont
  {Vexiau}}, \bibinfo {author} {\bibfnamefont {N.}~\bibnamefont
  {Bouloufa-Maafa}}, \bibinfo {author} {\bibfnamefont {G.}~\bibnamefont
  {Quemener}}, \bibinfo {author} {\bibfnamefont {O.}~\bibnamefont {Dulieu}},\
  and\ \bibinfo {author} {\bibfnamefont {D.}~\bibnamefont {Wang}},\ }\href
  {https://doi.org/10.1103/PhysRevLett.116.205303} {\bibfield  {journal}
  {\bibinfo  {journal} {PHYSICAL REVIEW LETTERS}\ }\textbf {\bibinfo {volume}
  {116}},\ \bibinfo {pages} {205303} (\bibinfo {year} {2016})}\BibitemShut
  {NoStop}%
\bibitem [{\citenamefont {Guo}\ \emph {et~al.}(2017)\citenamefont {Guo},
  \citenamefont {Vexiau}, \citenamefont {Zhu}, \citenamefont {Lu},
  \citenamefont {Bouloufa-Maafa}, \citenamefont {Dulieu},\ and\ \citenamefont
  {Wang}}]{MA2Na23Rb872017}%
  \BibitemOpen
  \bibfield  {author} {\bibinfo {author} {\bibfnamefont {M.}~\bibnamefont
  {Guo}}, \bibinfo {author} {\bibfnamefont {R.}~\bibnamefont {Vexiau}},
  \bibinfo {author} {\bibfnamefont {B.}~\bibnamefont {Zhu}}, \bibinfo {author}
  {\bibfnamefont {B.}~\bibnamefont {Lu}}, \bibinfo {author} {\bibfnamefont
  {N.}~\bibnamefont {Bouloufa-Maafa}}, \bibinfo {author} {\bibfnamefont
  {O.}~\bibnamefont {Dulieu}},\ and\ \bibinfo {author} {\bibfnamefont
  {D.}~\bibnamefont {Wang}},\ }\href
  {https://doi.org/10.1103/PhysRevA.96.052505} {\bibfield  {journal} {\bibinfo
  {journal} {PHYSICAL REVIEW A}\ }\textbf {\bibinfo {volume} {96}},\ \bibinfo
  {pages} {052505} (\bibinfo {year} {2017})}\BibitemShut {NoStop}%
\bibitem [{\citenamefont {Picard}\ \emph {et~al.}(2023)\citenamefont {Picard},
  \citenamefont {Zhang}, \citenamefont {Cairncross}, \citenamefont {Wang},
  \citenamefont {Patenotte}, \citenamefont {Park}, \citenamefont {Yu},
  \citenamefont {Liu}, \citenamefont {Hood}, \citenamefont
  {Gonz\'alez-F\'erez},\ and\ \citenamefont {Ni}}]{MANaCsNi2023}%
  \BibitemOpen
  \bibfield  {author} {\bibinfo {author} {\bibfnamefont {L.~R.~B.}\
  \bibnamefont {Picard}}, \bibinfo {author} {\bibfnamefont {J.~T.}\
  \bibnamefont {Zhang}}, \bibinfo {author} {\bibfnamefont {W.~B.}\ \bibnamefont
  {Cairncross}}, \bibinfo {author} {\bibfnamefont {K.}~\bibnamefont {Wang}},
  \bibinfo {author} {\bibfnamefont {G.~E.}\ \bibnamefont {Patenotte}}, \bibinfo
  {author} {\bibfnamefont {A.~J.}\ \bibnamefont {Park}}, \bibinfo {author}
  {\bibfnamefont {Y.}~\bibnamefont {Yu}}, \bibinfo {author} {\bibfnamefont
  {L.~R.}\ \bibnamefont {Liu}}, \bibinfo {author} {\bibfnamefont {J.~D.}\
  \bibnamefont {Hood}}, \bibinfo {author} {\bibfnamefont {R.}~\bibnamefont
  {Gonz\'alez-F\'erez}},\ and\ \bibinfo {author} {\bibfnamefont {K.-K.}\
  \bibnamefont {Ni}},\ }\href
  {https://doi.org/10.1103/PhysRevResearch.5.023149} {\bibfield  {journal}
  {\bibinfo  {journal} {Phys. Rev. Res.}\ }\textbf {\bibinfo {volume} {5}},\
  \bibinfo {pages} {023149} (\bibinfo {year} {2023})}\BibitemShut {NoStop}%
\bibitem [{\citenamefont {Stevenson}\ \emph {et~al.}(2023)\citenamefont
  {Stevenson}, \citenamefont {Lam}, \citenamefont {Bigagli}, \citenamefont
  {Warner}, \citenamefont {Yuan}, \citenamefont {Zhang},\ and\ \citenamefont
  {Will}}]{MANaCsStevenson2023}%
  \BibitemOpen
  \bibfield  {author} {\bibinfo {author} {\bibfnamefont {I.}~\bibnamefont
  {Stevenson}}, \bibinfo {author} {\bibfnamefont {A.~Z.}\ \bibnamefont {Lam}},
  \bibinfo {author} {\bibfnamefont {N.}~\bibnamefont {Bigagli}}, \bibinfo
  {author} {\bibfnamefont {C.}~\bibnamefont {Warner}}, \bibinfo {author}
  {\bibfnamefont {W.}~\bibnamefont {Yuan}}, \bibinfo {author} {\bibfnamefont
  {S.}~\bibnamefont {Zhang}},\ and\ \bibinfo {author} {\bibfnamefont
  {S.}~\bibnamefont {Will}},\ }\href
  {https://doi.org/10.1103/PhysRevLett.130.113002} {\bibfield  {journal}
  {\bibinfo  {journal} {PHYSICAL REVIEW LETTERS}\ }\textbf {\bibinfo {volume}
  {130}},\ \bibinfo {pages} {113002} (\bibinfo {year} {2023})}\BibitemShut
  {NoStop}%
\bibitem [{\citenamefont {Ni}\ \emph {et~al.}(2008)\citenamefont {Ni},
  \citenamefont {Ospelkaus}, \citenamefont {de~Miranda}, \citenamefont {Pe'er},
  \citenamefont {Neyenhuis}, \citenamefont {Zirbel}, \citenamefont
  {Kotochigova}, \citenamefont {Julienne}, \citenamefont {Jin},\ and\
  \citenamefont {Ye}}]{MAK40Rb872008}%
  \BibitemOpen
  \bibfield  {author} {\bibinfo {author} {\bibfnamefont {K.~K.}\ \bibnamefont
  {Ni}}, \bibinfo {author} {\bibfnamefont {S.}~\bibnamefont {Ospelkaus}},
  \bibinfo {author} {\bibfnamefont {M.~H.~G.}\ \bibnamefont {de~Miranda}},
  \bibinfo {author} {\bibfnamefont {A.}~\bibnamefont {Pe'er}}, \bibinfo
  {author} {\bibfnamefont {B.}~\bibnamefont {Neyenhuis}}, \bibinfo {author}
  {\bibfnamefont {J.~J.}\ \bibnamefont {Zirbel}}, \bibinfo {author}
  {\bibfnamefont {S.}~\bibnamefont {Kotochigova}}, \bibinfo {author}
  {\bibfnamefont {P.~S.}\ \bibnamefont {Julienne}}, \bibinfo {author}
  {\bibfnamefont {D.~S.}\ \bibnamefont {Jin}},\ and\ \bibinfo {author}
  {\bibfnamefont {J.}~\bibnamefont {Ye}},\ }\href
  {https://doi.org/10.1126/science.1163861} {\bibfield  {journal} {\bibinfo
  {journal} {SCIENCE}\ }\textbf {\bibinfo {volume} {322}},\ \bibinfo {pages}
  {231} (\bibinfo {year} {2008})}\BibitemShut {NoStop}%
\bibitem [{\citenamefont {Molony}\ \emph {et~al.}(2014)\citenamefont {Molony},
  \citenamefont {Gregory}, \citenamefont {Ji}, \citenamefont {Lu},
  \citenamefont {K\"oppinger}, \citenamefont {Le~Sueur}, \citenamefont
  {Blackley}, \citenamefont {Hutson},\ and\ \citenamefont
  {Cornish}}]{MARbCsCornish2014}%
  \BibitemOpen
  \bibfield  {author} {\bibinfo {author} {\bibfnamefont {P.~K.}\ \bibnamefont
  {Molony}}, \bibinfo {author} {\bibfnamefont {P.~D.}\ \bibnamefont {Gregory}},
  \bibinfo {author} {\bibfnamefont {Z.}~\bibnamefont {Ji}}, \bibinfo {author}
  {\bibfnamefont {B.}~\bibnamefont {Lu}}, \bibinfo {author} {\bibfnamefont
  {M.~P.}\ \bibnamefont {K\"oppinger}}, \bibinfo {author} {\bibfnamefont
  {C.~R.}\ \bibnamefont {Le~Sueur}}, \bibinfo {author} {\bibfnamefont {C.~L.}\
  \bibnamefont {Blackley}}, \bibinfo {author} {\bibfnamefont {J.~M.}\
  \bibnamefont {Hutson}},\ and\ \bibinfo {author} {\bibfnamefont {S.~L.}\
  \bibnamefont {Cornish}},\ }\href
  {https://doi.org/10.1103/PhysRevLett.113.255301} {\bibfield  {journal}
  {\bibinfo  {journal} {Phys. Rev. Lett.}\ }\textbf {\bibinfo {volume} {113}},\
  \bibinfo {pages} {255301} (\bibinfo {year} {2014})}\BibitemShut {NoStop}%
\bibitem [{\citenamefont {Takekoshi}\ \emph {et~al.}(2014)\citenamefont
  {Takekoshi}, \citenamefont {Reichsoellner}, \citenamefont {Schindewolf},
  \citenamefont {Hutson}, \citenamefont {Le~Sueur}, \citenamefont {Dulieu},
  \citenamefont {Ferlaino}, \citenamefont {Grimm},\ and\ \citenamefont
  {Naegerl}}]{MARb87Cs133Naegerl2014}%
  \BibitemOpen
  \bibfield  {author} {\bibinfo {author} {\bibfnamefont {T.}~\bibnamefont
  {Takekoshi}}, \bibinfo {author} {\bibfnamefont {L.}~\bibnamefont
  {Reichsoellner}}, \bibinfo {author} {\bibfnamefont {A.}~\bibnamefont
  {Schindewolf}}, \bibinfo {author} {\bibfnamefont {J.~M.}\ \bibnamefont
  {Hutson}}, \bibinfo {author} {\bibfnamefont {C.~R.}\ \bibnamefont
  {Le~Sueur}}, \bibinfo {author} {\bibfnamefont {O.}~\bibnamefont {Dulieu}},
  \bibinfo {author} {\bibfnamefont {F.}~\bibnamefont {Ferlaino}}, \bibinfo
  {author} {\bibfnamefont {R.}~\bibnamefont {Grimm}},\ and\ \bibinfo {author}
  {\bibfnamefont {H.-C.}\ \bibnamefont {Naegerl}},\ }\href
  {https://doi.org/10.1103/PhysRevLett.113.205301} {\bibfield  {journal}
  {\bibinfo  {journal} {PHYSICAL REVIEW LETTERS}\ }\textbf {\bibinfo {volume}
  {113}},\ \bibinfo {pages} {205301} (\bibinfo {year} {2014})}\BibitemShut
  {NoStop}%
\bibitem [{\citenamefont {Das}\ \emph {et~al.}(2023)\citenamefont {Das},
  \citenamefont {Gregory}, \citenamefont {Takekoshi}, \citenamefont {Fernley},
  \citenamefont {Landini}, \citenamefont {Hutson}, \citenamefont {Cornish},\
  and\ \citenamefont {Nägerl}}]{MARb87Cs1332023}%
  \BibitemOpen
  \bibfield  {author} {\bibinfo {author} {\bibfnamefont {A.}~\bibnamefont
  {Das}}, \bibinfo {author} {\bibfnamefont {P.~D.}\ \bibnamefont {Gregory}},
  \bibinfo {author} {\bibfnamefont {T.}~\bibnamefont {Takekoshi}}, \bibinfo
  {author} {\bibfnamefont {L.}~\bibnamefont {Fernley}}, \bibinfo {author}
  {\bibfnamefont {M.}~\bibnamefont {Landini}}, \bibinfo {author} {\bibfnamefont
  {J.~M.}\ \bibnamefont {Hutson}}, \bibinfo {author} {\bibfnamefont {S.~L.}\
  \bibnamefont {Cornish}},\ and\ \bibinfo {author} {\bibfnamefont {H.-C.}\
  \bibnamefont {Nägerl}},\ }\href
  {https://doi.org/10.21468/SciPostPhys.15.6.220} {\bibfield  {journal}
  {\bibinfo  {journal} {SciPost Phys.}\ }\textbf {\bibinfo {volume} {15}},\
  \bibinfo {pages} {220} (\bibinfo {year} {2023})}\BibitemShut {NoStop}%
\bibitem [{\citenamefont {Stevenson}(2018)}]{ICSThesis2018}%
  \BibitemOpen
  \bibfield  {author} {\bibinfo {author} {\bibfnamefont {I.~C.}\ \bibnamefont
  {Stevenson}},\ }\emph {\bibinfo {title} {Spectroscopy of Ultracold
  Lithium-Rubidium Molecules}},\ \href@noop {} {Ph.D. thesis},\ \bibinfo
  {school} {Purdue University} (\bibinfo {year} {2018})\BibitemShut {NoStop}%
\bibitem [{\citenamefont {Blasing}(2018)}]{BlasingThesis2018}%
  \BibitemOpen
  \bibfield  {author} {\bibinfo {author} {\bibfnamefont {D.}~\bibnamefont
  {Blasing}},\ }\emph {\bibinfo {title} {Photoassociation in 87RB BECS and in
  Ultracold 7LI85RB}},\ \href@noop {} {Ph.D. thesis},\ \bibinfo  {school}
  {Purdue University} (\bibinfo {year} {2018})\BibitemShut {NoStop}%
\bibitem [{\citenamefont {Li}\ \emph {et~al.}(2020)\citenamefont {Li},
  \citenamefont {Wang}, \citenamefont {Wang},\ and\ \citenamefont
  {Cong}}]{ZiangPRA2020}%
  \BibitemOpen
  \bibfield  {author} {\bibinfo {author} {\bibfnamefont {Z.-A.}\ \bibnamefont
  {Li}}, \bibinfo {author} {\bibfnamefont {Z.-W.}\ \bibnamefont {Wang}},
  \bibinfo {author} {\bibfnamefont {G.-R.}\ \bibnamefont {Wang}},\ and\
  \bibinfo {author} {\bibfnamefont {S.-L.}\ \bibnamefont {Cong}},\ }\href
  {https://doi.org/10.1103/PhysRevA.101.053409} {\bibfield  {journal} {\bibinfo
   {journal} {Phys. Rev. A}\ }\textbf {\bibinfo {volume} {101}},\ \bibinfo
  {pages} {053409} (\bibinfo {year} {2020})}\BibitemShut {NoStop}%
\bibitem [{\citenamefont {Valtolina}\ \emph {et~al.}(2020)\citenamefont
  {Valtolina}, \citenamefont {Matsuda}, \citenamefont {Tobias}, \citenamefont
  {Li}, \citenamefont {De~Marco},\ and\ \citenamefont
  {Ye}}]{ValtolinaNature2020}%
  \BibitemOpen
  \bibfield  {author} {\bibinfo {author} {\bibfnamefont {G.}~\bibnamefont
  {Valtolina}}, \bibinfo {author} {\bibfnamefont {K.}~\bibnamefont {Matsuda}},
  \bibinfo {author} {\bibfnamefont {W.~G.}\ \bibnamefont {Tobias}}, \bibinfo
  {author} {\bibfnamefont {J.-R.}\ \bibnamefont {Li}}, \bibinfo {author}
  {\bibfnamefont {L.}~\bibnamefont {De~Marco}},\ and\ \bibinfo {author}
  {\bibfnamefont {J.}~\bibnamefont {Ye}},\ }\href
  {https://doi.org/10.1038/s41586-020-2980-7} {\bibfield  {journal} {\bibinfo
  {journal} {Nature}\ }\textbf {\bibinfo {volume} {588}},\ \bibinfo {pages}
  {239} (\bibinfo {year} {2020})}\BibitemShut {NoStop}%
\bibitem [{\citenamefont {Schindewolf}\ \emph {et~al.}(2022)\citenamefont
  {Schindewolf}, \citenamefont {Bause}, \citenamefont {Chen}, \citenamefont
  {Duda}, \citenamefont {Karman}, \citenamefont {Bloch},\ and\ \citenamefont
  {Luo}}]{SchindewolfNature2022}%
  \BibitemOpen
  \bibfield  {author} {\bibinfo {author} {\bibfnamefont {A.}~\bibnamefont
  {Schindewolf}}, \bibinfo {author} {\bibfnamefont {R.}~\bibnamefont {Bause}},
  \bibinfo {author} {\bibfnamefont {X.-Y.}\ \bibnamefont {Chen}}, \bibinfo
  {author} {\bibfnamefont {M.}~\bibnamefont {Duda}}, \bibinfo {author}
  {\bibfnamefont {T.}~\bibnamefont {Karman}}, \bibinfo {author} {\bibfnamefont
  {I.}~\bibnamefont {Bloch}},\ and\ \bibinfo {author} {\bibfnamefont {X.-Y.}\
  \bibnamefont {Luo}},\ }\href {https://doi.org/10.1038/s41586-022-04900-0}
  {\bibfield  {journal} {\bibinfo  {journal} {Nature}\ }\textbf {\bibinfo
  {volume} {607}},\ \bibinfo {pages} {677} (\bibinfo {year}
  {2022})}\BibitemShut {NoStop}%
\bibitem [{\citenamefont {Deng}\ \emph {et~al.}(2023)\citenamefont {Deng},
  \citenamefont {Chen}, \citenamefont {Luo}, \citenamefont {Zhang},
  \citenamefont {Yi},\ and\ \citenamefont {Shi}}]{TaoShiPRL2022}%
  \BibitemOpen
  \bibfield  {author} {\bibinfo {author} {\bibfnamefont {F.}~\bibnamefont
  {Deng}}, \bibinfo {author} {\bibfnamefont {X.-Y.}\ \bibnamefont {Chen}},
  \bibinfo {author} {\bibfnamefont {X.-Y.}\ \bibnamefont {Luo}}, \bibinfo
  {author} {\bibfnamefont {W.}~\bibnamefont {Zhang}}, \bibinfo {author}
  {\bibfnamefont {S.}~\bibnamefont {Yi}},\ and\ \bibinfo {author}
  {\bibfnamefont {T.}~\bibnamefont {Shi}},\ }\href
  {https://doi.org/10.1103/PhysRevLett.130.183001} {\bibfield  {journal}
  {\bibinfo  {journal} {Phys. Rev. Lett.}\ }\textbf {\bibinfo {volume} {130}},\
  \bibinfo {pages} {183001} (\bibinfo {year} {2023})}\BibitemShut {NoStop}%
\bibitem [{\citenamefont {Zhang}\ \emph {et~al.}(2022)\citenamefont {Zhang},
  \citenamefont {Han}, \citenamefont {Cong},\ and\ \citenamefont
  {Shundalau}}]{FBZhangrong2022}%
  \BibitemOpen
  \bibfield  {author} {\bibinfo {author} {\bibfnamefont {R.}~\bibnamefont
  {Zhang}}, \bibinfo {author} {\bibfnamefont {Y.-C.}\ \bibnamefont {Han}},
  \bibinfo {author} {\bibfnamefont {S.-L.}\ \bibnamefont {Cong}},\ and\
  \bibinfo {author} {\bibfnamefont {M.~B.}\ \bibnamefont {Shundalau}},\ }\href
  {https://doi.org/10.1088/1674-1056/ac4cc3} {\bibfield  {journal} {\bibinfo
  {journal} {CHINESE PHYSICS B}\ }\textbf {\bibinfo {volume} {31}},\ \bibinfo
  {pages} {063402} (\bibinfo {year} {2022})}\BibitemShut {NoStop}%
\bibitem [{\citenamefont {Marzok}\ \emph {et~al.}(2009)\citenamefont {Marzok},
  \citenamefont {Deh}, \citenamefont {Zimmermann}, \citenamefont {Courteille},
  \citenamefont {Tiemann}, \citenamefont {Vanne},\ and\ \citenamefont
  {Saenz}}]{FBLi7Rb872009}%
  \BibitemOpen
  \bibfield  {author} {\bibinfo {author} {\bibfnamefont {C.}~\bibnamefont
  {Marzok}}, \bibinfo {author} {\bibfnamefont {B.}~\bibnamefont {Deh}},
  \bibinfo {author} {\bibfnamefont {C.}~\bibnamefont {Zimmermann}}, \bibinfo
  {author} {\bibfnamefont {P.~W.}\ \bibnamefont {Courteille}}, \bibinfo
  {author} {\bibfnamefont {E.}~\bibnamefont {Tiemann}}, \bibinfo {author}
  {\bibfnamefont {Y.~V.}\ \bibnamefont {Vanne}},\ and\ \bibinfo {author}
  {\bibfnamefont {A.}~\bibnamefont {Saenz}},\ }\href
  {https://doi.org/10.1103/PhysRevA.79.012717} {\bibfield  {journal} {\bibinfo
  {journal} {PHYSICAL REVIEW A}\ }\textbf {\bibinfo {volume} {79}},\ \bibinfo
  {pages} {012717} (\bibinfo {year} {2009})}\BibitemShut {NoStop}%
\bibitem [{\citenamefont {Ivanova}\ \emph {et~al.}(2011)\citenamefont
  {Ivanova}, \citenamefont {Stein}, \citenamefont {Pashov}, \citenamefont
  {Knoeckel},\ and\ \citenamefont {Tiemann}}]{TiemannIPAX1Sa3S2011}%
  \BibitemOpen
  \bibfield  {author} {\bibinfo {author} {\bibfnamefont {M.}~\bibnamefont
  {Ivanova}}, \bibinfo {author} {\bibfnamefont {A.}~\bibnamefont {Stein}},
  \bibinfo {author} {\bibfnamefont {A.}~\bibnamefont {Pashov}}, \bibinfo
  {author} {\bibfnamefont {H.}~\bibnamefont {Knoeckel}},\ and\ \bibinfo
  {author} {\bibfnamefont {E.}~\bibnamefont {Tiemann}},\ }\href
  {https://doi.org/10.1063/1.3524312} {\bibfield  {journal} {\bibinfo
  {journal} {JOURNAL OF CHEMICAL PHYSICS}\ }\textbf {\bibinfo {volume} {134}},\
  \bibinfo {pages} {024321} (\bibinfo {year} {2011})}\BibitemShut {NoStop}%
\bibitem [{\citenamefont {Ivanova}\ \emph {et~al.}(2013)\citenamefont
  {Ivanova}, \citenamefont {Stein}, \citenamefont {Pashov}, \citenamefont
  {Knoeckel},\ and\ \citenamefont {Tiemann}}]{TiemannLEVELB1PiD1Pi2013}%
  \BibitemOpen
  \bibfield  {author} {\bibinfo {author} {\bibfnamefont {M.}~\bibnamefont
  {Ivanova}}, \bibinfo {author} {\bibfnamefont {A.}~\bibnamefont {Stein}},
  \bibinfo {author} {\bibfnamefont {A.}~\bibnamefont {Pashov}}, \bibinfo
  {author} {\bibfnamefont {H.}~\bibnamefont {Knoeckel}},\ and\ \bibinfo
  {author} {\bibfnamefont {E.}~\bibnamefont {Tiemann}},\ }\href
  {https://doi.org/10.1063/1.4793315} {\bibfield  {journal} {\bibinfo
  {journal} {JOURNAL OF CHEMICAL PHYSICS}\ }\textbf {\bibinfo {volume} {138}},\
  \bibinfo {pages} {094315} (\bibinfo {year} {2013})}\BibitemShut {NoStop}%
\bibitem [{\citenamefont {Dutta}\ \emph {et~al.}(2011)\citenamefont {Dutta},
  \citenamefont {Altaf}, \citenamefont {Elliott},\ and\ \citenamefont
  {Chen}}]{LiRbLEVEL1B1Pi2011}%
  \BibitemOpen
  \bibfield  {author} {\bibinfo {author} {\bibfnamefont {S.}~\bibnamefont
  {Dutta}}, \bibinfo {author} {\bibfnamefont {A.}~\bibnamefont {Altaf}},
  \bibinfo {author} {\bibfnamefont {D.~S.}\ \bibnamefont {Elliott}},\ and\
  \bibinfo {author} {\bibfnamefont {Y.~P.}\ \bibnamefont {Chen}},\ }\href
  {https://doi.org/10.1016/j.cplett.2011.05.059} {\bibfield  {journal}
  {\bibinfo  {journal} {CHEMICAL PHYSICS LETTERS}\ }\textbf {\bibinfo {volume}
  {511}},\ \bibinfo {pages} {7} (\bibinfo {year} {2011})}\BibitemShut {NoStop}%
\bibitem [{\citenamefont {Dutta}\ \emph
  {et~al.}(2014{\natexlab{a}})\citenamefont {Dutta}, \citenamefont {Elliott},\
  and\ \citenamefont {Chen}}]{add1Dutta2013}%
  \BibitemOpen
  \bibfield  {author} {\bibinfo {author} {\bibfnamefont {S.}~\bibnamefont
  {Dutta}}, \bibinfo {author} {\bibfnamefont {D.~S.}\ \bibnamefont {Elliott}},\
  and\ \bibinfo {author} {\bibfnamefont {Y.~P.}\ \bibnamefont {Chen}},\ }\href
  {https://doi.org/10.1209/0295-5075/104/63001} {\bibfield  {journal} {\bibinfo
   {journal} {Europhysics Letters}\ }\textbf {\bibinfo {volume} {104}},\
  \bibinfo {pages} {63001} (\bibinfo {year} {2014}{\natexlab{a}})}\BibitemShut
  {NoStop}%
\bibitem [{\citenamefont {Dutta}\ \emph
  {et~al.}(2014{\natexlab{b}})\citenamefont {Dutta}, \citenamefont {Lorenz},
  \citenamefont {Altaf}, \citenamefont {Elliott},\ and\ \citenamefont
  {Chen}}]{add2Dutta2014}%
  \BibitemOpen
  \bibfield  {author} {\bibinfo {author} {\bibfnamefont {S.}~\bibnamefont
  {Dutta}}, \bibinfo {author} {\bibfnamefont {J.}~\bibnamefont {Lorenz}},
  \bibinfo {author} {\bibfnamefont {A.}~\bibnamefont {Altaf}}, \bibinfo
  {author} {\bibfnamefont {D.~S.}\ \bibnamefont {Elliott}},\ and\ \bibinfo
  {author} {\bibfnamefont {Y.~P.}\ \bibnamefont {Chen}},\ }\href
  {https://doi.org/10.1103/PhysRevA.89.020702} {\bibfield  {journal} {\bibinfo
  {journal} {Phys. Rev. A}\ }\textbf {\bibinfo {volume} {89}},\ \bibinfo
  {pages} {020702} (\bibinfo {year} {2014}{\natexlab{b}})}\BibitemShut
  {NoStop}%
\bibitem [{\citenamefont {Lorenz}\ \emph {et~al.}(2014)\citenamefont {Lorenz},
  \citenamefont {Altaf}, \citenamefont {Dutta}, \citenamefont {Chen},\ and\
  \citenamefont {Elliott}}]{LiRbLEVEL3B1PiD1Pi2014}%
  \BibitemOpen
  \bibfield  {author} {\bibinfo {author} {\bibfnamefont {J.}~\bibnamefont
  {Lorenz}}, \bibinfo {author} {\bibfnamefont {A.}~\bibnamefont {Altaf}},
  \bibinfo {author} {\bibfnamefont {S.}~\bibnamefont {Dutta}}, \bibinfo
  {author} {\bibfnamefont {Y.~P.}\ \bibnamefont {Chen}},\ and\ \bibinfo
  {author} {\bibfnamefont {D.~S.}\ \bibnamefont {Elliott}},\ }\href
  {https://doi.org/10.1103/PhysRevA.90.062513} {\bibfield  {journal} {\bibinfo
  {journal} {PHYSICAL REVIEW A}\ }\textbf {\bibinfo {volume} {90}},\ \bibinfo
  {pages} {062513} (\bibinfo {year} {2014})}\BibitemShut {NoStop}%
\bibitem [{\citenamefont {Altaf}\ \emph {et~al.}(2015)\citenamefont {Altaf},
  \citenamefont {Dutta}, \citenamefont {Lorenz}, \citenamefont {Pérez-Ríos},
  \citenamefont {Chen},\ and\ \citenamefont
  {Elliott}}]{LiRbLEVELa3Sig+f3Pig3Sig+2015}%
  \BibitemOpen
  \bibfield  {author} {\bibinfo {author} {\bibfnamefont {A.}~\bibnamefont
  {Altaf}}, \bibinfo {author} {\bibfnamefont {S.}~\bibnamefont {Dutta}},
  \bibinfo {author} {\bibfnamefont {J.}~\bibnamefont {Lorenz}}, \bibinfo
  {author} {\bibfnamefont {J.}~\bibnamefont {Pérez-Ríos}}, \bibinfo {author}
  {\bibfnamefont {Y.~P.}\ \bibnamefont {Chen}},\ and\ \bibinfo {author}
  {\bibfnamefont {D.~S.}\ \bibnamefont {Elliott}},\ }\href
  {https://doi.org/10.1063/1.4914917} {\bibfield  {journal} {\bibinfo
  {journal} {JOURNAL OF CHEMICAL PHYSICS}\ }\textbf {\bibinfo {volume} {142}},\
  \bibinfo {pages} {114310} (\bibinfo {year} {2015})}\BibitemShut {NoStop}%
\bibitem [{\citenamefont {Blasing}\ \emph {et~al.}(2016)\citenamefont
  {Blasing}, \citenamefont {Stevenson}, \citenamefont {Pérez-Ríos},
  \citenamefont {Elliott},\ and\ \citenamefont {Chen}}]{PALiRba3Sig+2016}%
  \BibitemOpen
  \bibfield  {author} {\bibinfo {author} {\bibfnamefont {D.~B.}\ \bibnamefont
  {Blasing}}, \bibinfo {author} {\bibfnamefont {I.~C.}\ \bibnamefont
  {Stevenson}}, \bibinfo {author} {\bibfnamefont {J.}~\bibnamefont
  {Pérez-Ríos}}, \bibinfo {author} {\bibfnamefont {D.~S.}\ \bibnamefont
  {Elliott}},\ and\ \bibinfo {author} {\bibfnamefont {Y.~P.}\ \bibnamefont
  {Chen}},\ }\href {https://doi.org/10.1103/PhysRevA.94.062504} {\bibfield
  {journal} {\bibinfo  {journal} {PHYSICAL REVIEW A}\ }\textbf {\bibinfo
  {volume} {94}},\ \bibinfo {pages} {062504} (\bibinfo {year}
  {2016})}\BibitemShut {NoStop}%
\bibitem [{\citenamefont {Stevenson}\ \emph
  {et~al.}(2016{\natexlab{a}})\citenamefont {Stevenson}, \citenamefont
  {Blasing}, \citenamefont {Chen},\ and\ \citenamefont
  {Elliott}}]{LiRbLEVELA1b3C12016}%
  \BibitemOpen
  \bibfield  {author} {\bibinfo {author} {\bibfnamefont {I.~C.}\ \bibnamefont
  {Stevenson}}, \bibinfo {author} {\bibfnamefont {D.~B.}\ \bibnamefont
  {Blasing}}, \bibinfo {author} {\bibfnamefont {Y.~P.}\ \bibnamefont {Chen}},\
  and\ \bibinfo {author} {\bibfnamefont {D.~S.}\ \bibnamefont {Elliott}},\
  }\href {https://doi.org/10.1103/PhysRevA.94.062503} {\bibfield  {journal}
  {\bibinfo  {journal} {PHYSICAL REVIEW A}\ }\textbf {\bibinfo {volume} {94}},\
  \bibinfo {pages} {062503} (\bibinfo {year} {2016}{\natexlab{a}})}\BibitemShut
  {NoStop}%
\bibitem [{\citenamefont {Stevenson}\ \emph
  {et~al.}(2016{\natexlab{b}})\citenamefont {Stevenson}, \citenamefont
  {Blasing}, \citenamefont {Altaf}, \citenamefont {Chen},\ and\ \citenamefont
  {Elliott}}]{LiRbLEVELd3Pi2016}%
  \BibitemOpen
  \bibfield  {author} {\bibinfo {author} {\bibfnamefont {I.~C.}\ \bibnamefont
  {Stevenson}}, \bibinfo {author} {\bibfnamefont {D.~B.}\ \bibnamefont
  {Blasing}}, \bibinfo {author} {\bibfnamefont {A.}~\bibnamefont {Altaf}},
  \bibinfo {author} {\bibfnamefont {Y.~P.}\ \bibnamefont {Chen}},\ and\
  \bibinfo {author} {\bibfnamefont {D.~S.}\ \bibnamefont {Elliott}},\ }\href
  {https://doi.org/10.1063/1.4964655} {\bibfield  {journal} {\bibinfo
  {journal} {JOURNAL OF CHEMICAL PHYSICS}\ }\textbf {\bibinfo {volume} {145}},\
  \bibinfo {pages} {224301} (\bibinfo {year} {2016}{\natexlab{b}})}\BibitemShut
  {NoStop}%
\bibitem [{\citenamefont {Stevenson}\ \emph
  {et~al.}(2016{\natexlab{c}})\citenamefont {Stevenson}, \citenamefont
  {Blasing}, \citenamefont {Chen},\ and\ \citenamefont
  {Elliott}}]{PALi7Rb852016}%
  \BibitemOpen
  \bibfield  {author} {\bibinfo {author} {\bibfnamefont {I.~C.}\ \bibnamefont
  {Stevenson}}, \bibinfo {author} {\bibfnamefont {D.~B.}\ \bibnamefont
  {Blasing}}, \bibinfo {author} {\bibfnamefont {Y.~P.}\ \bibnamefont {Chen}},\
  and\ \bibinfo {author} {\bibfnamefont {D.~S.}\ \bibnamefont {Elliott}},\
  }\href {https://doi.org/10.1103/PhysRevA.94.062510} {\bibfield  {journal}
  {\bibinfo  {journal} {PHYSICAL REVIEW A}\ }\textbf {\bibinfo {volume} {94}},\
  \bibinfo {pages} {062510} (\bibinfo {year} {2016}{\natexlab{c}})}\BibitemShut
  {NoStop}%
\bibitem [{\citenamefont {Dutta}\ \emph {et~al.}(2017)\citenamefont {Dutta},
  \citenamefont {P\'erez-R\'{\i}os}, \citenamefont {Elliott},\ and\
  \citenamefont {Chen}}]{add3Dutta2017}%
  \BibitemOpen
  \bibfield  {author} {\bibinfo {author} {\bibfnamefont {S.}~\bibnamefont
  {Dutta}}, \bibinfo {author} {\bibfnamefont {J.}~\bibnamefont
  {P\'erez-R\'{\i}os}}, \bibinfo {author} {\bibfnamefont {D.~S.}\ \bibnamefont
  {Elliott}},\ and\ \bibinfo {author} {\bibfnamefont {Y.~P.}\ \bibnamefont
  {Chen}},\ }\href {https://doi.org/10.1103/PhysRevA.95.013405} {\bibfield
  {journal} {\bibinfo  {journal} {Phys. Rev. A}\ }\textbf {\bibinfo {volume}
  {95}},\ \bibinfo {pages} {013405} (\bibinfo {year} {2017})}\BibitemShut
  {NoStop}%
\bibitem [{\citenamefont {Korek}\ \emph {et~al.}(2009)\citenamefont {Korek},
  \citenamefont {Younes},\ and\ \citenamefont {Al-Shawa}}]{abPECLB2009}%
  \BibitemOpen
  \bibfield  {author} {\bibinfo {author} {\bibfnamefont {M.}~\bibnamefont
  {Korek}}, \bibinfo {author} {\bibfnamefont {G.}~\bibnamefont {Younes}},\ and\
  \bibinfo {author} {\bibfnamefont {S.}~\bibnamefont {Al-Shawa}},\ }\href
  {https://doi.org/10.1016/j.theochem.2008.12.006} {\bibfield  {journal}
  {\bibinfo  {journal} {JOURNAL OF MOLECULAR STRUCTURE-THEOCHEM}\ }\textbf
  {\bibinfo {volume} {899}},\ \bibinfo {pages} {25} (\bibinfo {year}
  {2009})}\BibitemShut {NoStop}%
\bibitem [{\citenamefont {Jendoubi}\ \emph {et~al.}(2012)\citenamefont
  {Jendoubi}, \citenamefont {Berriche}, \citenamefont {Ben~Ouada},\ and\
  \citenamefont {Gadea}}]{abTDM2012}%
  \BibitemOpen
  \bibfield  {author} {\bibinfo {author} {\bibfnamefont {I.}~\bibnamefont
  {Jendoubi}}, \bibinfo {author} {\bibfnamefont {H.}~\bibnamefont {Berriche}},
  \bibinfo {author} {\bibfnamefont {H.}~\bibnamefont {Ben~Ouada}},\ and\
  \bibinfo {author} {\bibfnamefont {F.~X.}\ \bibnamefont {Gadea}},\ }\href
  {https://doi.org/10.1021/jp209106w} {\bibfield  {journal} {\bibinfo
  {journal} {JOURNAL OF PHYSICAL CHEMISTRY A}\ }\textbf {\bibinfo {volume}
  {116}},\ \bibinfo {pages} {2945} (\bibinfo {year} {2012})}\BibitemShut
  {NoStop}%
\bibitem [{\citenamefont {You}\ \emph {et~al.}(2016)\citenamefont {You},
  \citenamefont {Yang}, \citenamefont {Zhang}, \citenamefont {Wang},
  \citenamefont {Ma},\ and\ \citenamefont {Liu}}]{abPECTDMCN2016}%
  \BibitemOpen
  \bibfield  {author} {\bibinfo {author} {\bibfnamefont {Y.}~\bibnamefont
  {You}}, \bibinfo {author} {\bibfnamefont {C.-L.}\ \bibnamefont {Yang}},
  \bibinfo {author} {\bibfnamefont {Q.-Q.}\ \bibnamefont {Zhang}}, \bibinfo
  {author} {\bibfnamefont {M.-S.}\ \bibnamefont {Wang}}, \bibinfo {author}
  {\bibfnamefont {X.-G.}\ \bibnamefont {Ma}},\ and\ \bibinfo {author}
  {\bibfnamefont {W.-W.}\ \bibnamefont {Liu}},\ }\href
  {https://doi.org/10.1039/c6cp01618a} {\bibfield  {journal} {\bibinfo
  {journal} {PHYSICAL CHEMISTRY CHEMICAL PHYSICS}\ }\textbf {\bibinfo {volume}
  {18}},\ \bibinfo {pages} {19838} (\bibinfo {year} {2016})}\BibitemShut
  {NoStop}%
\bibitem [{\citenamefont {Bormotova}\ \emph {et~al.}(2018)\citenamefont
  {Bormotova}, \citenamefont {Kozlov}, \citenamefont {Pazyuk},\ and\
  \citenamefont {Stolyarov}}]{abTDMPCCP2018}%
  \BibitemOpen
  \bibfield  {author} {\bibinfo {author} {\bibfnamefont {E.~A.}\ \bibnamefont
  {Bormotova}}, \bibinfo {author} {\bibfnamefont {S.~V.}\ \bibnamefont
  {Kozlov}}, \bibinfo {author} {\bibfnamefont {E.~A.}\ \bibnamefont {Pazyuk}},\
  and\ \bibinfo {author} {\bibfnamefont {A.~V.}\ \bibnamefont {Stolyarov}},\
  }\href {https://doi.org/10.1039/C7CP05548J} {\bibfield  {journal} {\bibinfo
  {journal} {Phys. Chem. Chem. Phys.}\ }\textbf {\bibinfo {volume} {20}},\
  \bibinfo {pages} {1889} (\bibinfo {year} {2018})}\BibitemShut {NoStop}%
\bibitem [{\citenamefont {Bormotova}\ \emph {et~al.}(2019)\citenamefont
  {Bormotova}, \citenamefont {Kozlov}, \citenamefont {Pazyuk}, \citenamefont
  {Stolyarov}, \citenamefont {Skomorowski}, \citenamefont {Majewska},\ and\
  \citenamefont {Moszynski}}]{abSOCPRA2019}%
  \BibitemOpen
  \bibfield  {author} {\bibinfo {author} {\bibfnamefont {E.~A.}\ \bibnamefont
  {Bormotova}}, \bibinfo {author} {\bibfnamefont {S.~V.}\ \bibnamefont
  {Kozlov}}, \bibinfo {author} {\bibfnamefont {E.~A.}\ \bibnamefont {Pazyuk}},
  \bibinfo {author} {\bibfnamefont {A.~V.}\ \bibnamefont {Stolyarov}}, \bibinfo
  {author} {\bibfnamefont {W.}~\bibnamefont {Skomorowski}}, \bibinfo {author}
  {\bibfnamefont {I.}~\bibnamefont {Majewska}},\ and\ \bibinfo {author}
  {\bibfnamefont {R.}~\bibnamefont {Moszynski}},\ }\href
  {https://doi.org/10.1103/PhysRevA.99.012507} {\bibfield  {journal} {\bibinfo
  {journal} {PHYSICAL REVIEW A}\ }\textbf {\bibinfo {volume} {99}},\ \bibinfo
  {pages} {012507} (\bibinfo {year} {2019})}\BibitemShut {NoStop}%
\bibitem [{\citenamefont {Kozlov}\ \emph {et~al.}(2020)\citenamefont {Kozlov},
  \citenamefont {Bormotova}, \citenamefont {Medvedev}, \citenamefont {Pazyuk},
  \citenamefont {Stolyarov},\ and\ \citenamefont
  {Zaitsevskii}}]{abPECSOCPCCP2020}%
  \BibitemOpen
  \bibfield  {author} {\bibinfo {author} {\bibfnamefont {S.}~\bibnamefont
  {Kozlov}, \bibfnamefont {V}}, \bibinfo {author} {\bibfnamefont {E.~A.}\
  \bibnamefont {Bormotova}}, \bibinfo {author} {\bibfnamefont {A.~A.}\
  \bibnamefont {Medvedev}}, \bibinfo {author} {\bibfnamefont {E.~A.}\
  \bibnamefont {Pazyuk}}, \bibinfo {author} {\bibfnamefont {A.}~\bibnamefont
  {Stolyarov}, \bibfnamefont {V}},\ and\ \bibinfo {author} {\bibfnamefont
  {A.}~\bibnamefont {Zaitsevskii}},\ }\href
  {https://doi.org/10.1039/c9cp06421d} {\bibfield  {journal} {\bibinfo
  {journal} {PHYSICAL CHEMISTRY CHEMICAL PHYSICS}\ }\textbf {\bibinfo {volume}
  {22}},\ \bibinfo {pages} {2295} (\bibinfo {year} {2020})}\BibitemShut
  {NoStop}%
\bibitem [{\citenamefont {Lefebvre-Brion}\ and\ \citenamefont
  {Field}(2004)}]{BookSpec&Dyna2004}%
  \BibitemOpen
  \bibfield  {author} {\bibinfo {author} {\bibfnamefont {H.}~\bibnamefont
  {Lefebvre-Brion}}\ and\ \bibinfo {author} {\bibfnamefont {R.~W.}\
  \bibnamefont {Field}},\ }\href
  {https://doi.org/10.1016/B978-0-12-441455-6.X5000-8,} {\emph {\bibinfo
  {title} {The Spectra and Dynamics of Diatomic Molecules}}}\ (\bibinfo
  {publisher} {Elsevier Inc.},\ \bibinfo {year} {2004})\BibitemShut {NoStop}%
\bibitem [{\citenamefont {Tamanis}\ \emph {et~al.}(2002)\citenamefont
  {Tamanis}, \citenamefont {Ferber}, \citenamefont {Zaitsevskii}, \citenamefont
  {Pazyuk}, \citenamefont {Stolyarov}, \citenamefont {Chen}, \citenamefont
  {Qi}, \citenamefont {Wang},\ and\ \citenamefont {Stwalley}}]{DP0NaRb2002}%
  \BibitemOpen
  \bibfield  {author} {\bibinfo {author} {\bibfnamefont {M.}~\bibnamefont
  {Tamanis}}, \bibinfo {author} {\bibfnamefont {R.}~\bibnamefont {Ferber}},
  \bibinfo {author} {\bibfnamefont {A.}~\bibnamefont {Zaitsevskii}}, \bibinfo
  {author} {\bibfnamefont {E.}~\bibnamefont {Pazyuk}}, \bibinfo {author}
  {\bibfnamefont {A.}~\bibnamefont {Stolyarov}}, \bibinfo {author}
  {\bibfnamefont {H.}~\bibnamefont {Chen}}, \bibinfo {author} {\bibfnamefont
  {J.}~\bibnamefont {Qi}}, \bibinfo {author} {\bibfnamefont {H.}~\bibnamefont
  {Wang}},\ and\ \bibinfo {author} {\bibfnamefont {W.}~\bibnamefont
  {Stwalley}},\ }\href {https://doi.org/10.1063/1.1505442} {\bibfield
  {journal} {\bibinfo  {journal} {JOURNAL OF CHEMICAL PHYSICS}\ }\textbf
  {\bibinfo {volume} {117}},\ \bibinfo {pages} {7980} (\bibinfo {year}
  {2002})}\BibitemShut {NoStop}%
\bibitem [{\citenamefont {Docenko}\ \emph {et~al.}(2007)\citenamefont
  {Docenko}, \citenamefont {Tamanis}, \citenamefont {Ferber}, \citenamefont
  {Pazyuk}, \citenamefont {Zaitsevskii}, \citenamefont {Stolyarov},
  \citenamefont {Pashov}, \citenamefont {Knoeckel},\ and\ \citenamefont
  {Tiemann}}]{DPNaRb2007}%
  \BibitemOpen
  \bibfield  {author} {\bibinfo {author} {\bibfnamefont {O.}~\bibnamefont
  {Docenko}}, \bibinfo {author} {\bibfnamefont {M.}~\bibnamefont {Tamanis}},
  \bibinfo {author} {\bibfnamefont {R.}~\bibnamefont {Ferber}}, \bibinfo
  {author} {\bibfnamefont {E.~A.}\ \bibnamefont {Pazyuk}}, \bibinfo {author}
  {\bibfnamefont {A.}~\bibnamefont {Zaitsevskii}}, \bibinfo {author}
  {\bibfnamefont {A.~V.}\ \bibnamefont {Stolyarov}}, \bibinfo {author}
  {\bibfnamefont {A.}~\bibnamefont {Pashov}}, \bibinfo {author} {\bibfnamefont
  {H.}~\bibnamefont {Knoeckel}},\ and\ \bibinfo {author} {\bibfnamefont
  {E.}~\bibnamefont {Tiemann}},\ }\href
  {https://doi.org/10.1103/PhysRevA.75.042503} {\bibfield  {journal} {\bibinfo
  {journal} {PHYSICAL REVIEW A}\ }\textbf {\bibinfo {volume} {75}},\ \bibinfo
  {pages} {042503} (\bibinfo {year} {2007})}\BibitemShut {NoStop}%
\bibitem [{\citenamefont {Bergeman}\ \emph {et~al.}(2003)\citenamefont
  {Bergeman}, \citenamefont {Fellows}, \citenamefont {Gutterres},\ and\
  \citenamefont {Amiot}}]{DP0RbCs2003}%
  \BibitemOpen
  \bibfield  {author} {\bibinfo {author} {\bibfnamefont {T.}~\bibnamefont
  {Bergeman}}, \bibinfo {author} {\bibfnamefont {C.}~\bibnamefont {Fellows}},
  \bibinfo {author} {\bibfnamefont {R.}~\bibnamefont {Gutterres}},\ and\
  \bibinfo {author} {\bibfnamefont {C.}~\bibnamefont {Amiot}},\ }\href
  {https://doi.org/10.1103/PhysRevA.67.050501} {\bibfield  {journal} {\bibinfo
  {journal} {PHYSICAL REVIEW A}\ }\textbf {\bibinfo {volume} {67}},\ \bibinfo
  {pages} {050501} (\bibinfo {year} {2003})}\BibitemShut {NoStop}%
\bibitem [{\citenamefont {Docenko}\ \emph {et~al.}(2010)\citenamefont
  {Docenko}, \citenamefont {Tamanis}, \citenamefont {Ferber}, \citenamefont
  {Bergeman}, \citenamefont {Kotochigova}, \citenamefont {Stolyarov},
  \citenamefont {Nogueira},\ and\ \citenamefont {Fellows}}]{DP1RbCs2010}%
  \BibitemOpen
  \bibfield  {author} {\bibinfo {author} {\bibfnamefont {O.}~\bibnamefont
  {Docenko}}, \bibinfo {author} {\bibfnamefont {M.}~\bibnamefont {Tamanis}},
  \bibinfo {author} {\bibfnamefont {R.}~\bibnamefont {Ferber}}, \bibinfo
  {author} {\bibfnamefont {T.}~\bibnamefont {Bergeman}}, \bibinfo {author}
  {\bibfnamefont {S.}~\bibnamefont {Kotochigova}}, \bibinfo {author}
  {\bibfnamefont {A.~V.}\ \bibnamefont {Stolyarov}}, \bibinfo {author}
  {\bibfnamefont {A.~d.~F.}\ \bibnamefont {Nogueira}},\ and\ \bibinfo {author}
  {\bibfnamefont {C.~E.}\ \bibnamefont {Fellows}},\ }\href
  {https://doi.org/10.1103/PhysRevA.81.042511} {\bibfield  {journal} {\bibinfo
  {journal} {PHYSICAL REVIEW A}\ }\textbf {\bibinfo {volume} {81}},\ \bibinfo
  {pages} {042511} (\bibinfo {year} {2010})}\BibitemShut {NoStop}%
\bibitem [{\citenamefont {Kruzins}\ \emph {et~al.}(2014)\citenamefont
  {Kruzins}, \citenamefont {Alps}, \citenamefont {Docenko}, \citenamefont
  {Klincare}, \citenamefont {Tamanis}, \citenamefont {Ferber}, \citenamefont
  {Pazyuk},\ and\ \citenamefont {Stolyarov}}]{DP2RbCs2014}%
  \BibitemOpen
  \bibfield  {author} {\bibinfo {author} {\bibfnamefont {A.}~\bibnamefont
  {Kruzins}}, \bibinfo {author} {\bibfnamefont {K.}~\bibnamefont {Alps}},
  \bibinfo {author} {\bibfnamefont {O.}~\bibnamefont {Docenko}}, \bibinfo
  {author} {\bibfnamefont {I.}~\bibnamefont {Klincare}}, \bibinfo {author}
  {\bibfnamefont {M.}~\bibnamefont {Tamanis}}, \bibinfo {author} {\bibfnamefont
  {R.}~\bibnamefont {Ferber}}, \bibinfo {author} {\bibfnamefont {E.~A.}\
  \bibnamefont {Pazyuk}},\ and\ \bibinfo {author} {\bibfnamefont {A.~V.}\
  \bibnamefont {Stolyarov}},\ }\href {https://doi.org/10.1063/1.4901327}
  {\bibfield  {journal} {\bibinfo  {journal} {JOURNAL OF CHEMICAL PHYSICS}\
  }\textbf {\bibinfo {volume} {141}},\ \bibinfo {pages} {184309} (\bibinfo
  {year} {2014})}\BibitemShut {NoStop}%
\bibitem [{\citenamefont {Qi}\ \emph {et~al.}(2007)\citenamefont {Qi},
  \citenamefont {Bai}, \citenamefont {Ahmed}, \citenamefont {Lyyra},
  \citenamefont {Kotochigova}, \citenamefont {Ross}, \citenamefont {Effantin},
  \citenamefont {Zalicki}, \citenamefont {Vigue}, \citenamefont {Chawla},
  \citenamefont {Field}, \citenamefont {Whang}, \citenamefont {Stwalley},
  \citenamefont {Knoeckel}, \citenamefont {Tiemann}, \citenamefont {Shang},
  \citenamefont {Li},\ and\ \citenamefont {Bergeman}}]{DPNaNa2007}%
  \BibitemOpen
  \bibfield  {author} {\bibinfo {author} {\bibfnamefont {P.}~\bibnamefont
  {Qi}}, \bibinfo {author} {\bibfnamefont {J.}~\bibnamefont {Bai}}, \bibinfo
  {author} {\bibfnamefont {E.}~\bibnamefont {Ahmed}}, \bibinfo {author}
  {\bibfnamefont {A.~M.}\ \bibnamefont {Lyyra}}, \bibinfo {author}
  {\bibfnamefont {S.}~\bibnamefont {Kotochigova}}, \bibinfo {author}
  {\bibfnamefont {A.~J.}\ \bibnamefont {Ross}}, \bibinfo {author}
  {\bibfnamefont {C.}~\bibnamefont {Effantin}}, \bibinfo {author}
  {\bibfnamefont {P.}~\bibnamefont {Zalicki}}, \bibinfo {author} {\bibfnamefont
  {J.}~\bibnamefont {Vigue}}, \bibinfo {author} {\bibfnamefont
  {G.}~\bibnamefont {Chawla}}, \bibinfo {author} {\bibfnamefont {R.~W.}\
  \bibnamefont {Field}}, \bibinfo {author} {\bibfnamefont {T.-J.}\ \bibnamefont
  {Whang}}, \bibinfo {author} {\bibfnamefont {W.~C.}\ \bibnamefont {Stwalley}},
  \bibinfo {author} {\bibfnamefont {H.}~\bibnamefont {Knoeckel}}, \bibinfo
  {author} {\bibfnamefont {E.}~\bibnamefont {Tiemann}}, \bibinfo {author}
  {\bibfnamefont {J.}~\bibnamefont {Shang}}, \bibinfo {author} {\bibfnamefont
  {L.}~\bibnamefont {Li}},\ and\ \bibinfo {author} {\bibfnamefont
  {T.}~\bibnamefont {Bergeman}},\ }\href {https://doi.org/10.1063/1.2747595}
  {\bibfield  {journal} {\bibinfo  {journal} {JOURNAL OF CHEMICAL PHYSICS}\
  }\textbf {\bibinfo {volume} {127}},\ \bibinfo {pages} {044301} (\bibinfo
  {year} {2007})}\BibitemShut {NoStop}%
\bibitem [{\citenamefont {Bai}\ \emph {et~al.}(2011)\citenamefont {Bai},
  \citenamefont {Ahmed}, \citenamefont {Beser}, \citenamefont {Guan},
  \citenamefont {Kotochigova}, \citenamefont {Lyyra}, \citenamefont {Ashman},
  \citenamefont {Wolfe}, \citenamefont {Huennekens}, \citenamefont {Xie},
  \citenamefont {Li}, \citenamefont {Li}, \citenamefont {Tamanis},
  \citenamefont {Ferber}, \citenamefont {Drozdova}, \citenamefont {Pazyuk},
  \citenamefont {Stolyarov}, \citenamefont {Danzl}, \citenamefont {Naegerl},
  \citenamefont {Bouloufa}, \citenamefont {Dulieu}, \citenamefont {Amiot},
  \citenamefont {Salami},\ and\ \citenamefont {Bergeman}}]{DP1CsCs2011}%
  \BibitemOpen
  \bibfield  {author} {\bibinfo {author} {\bibfnamefont {J.}~\bibnamefont
  {Bai}}, \bibinfo {author} {\bibfnamefont {E.~H.}\ \bibnamefont {Ahmed}},
  \bibinfo {author} {\bibfnamefont {B.}~\bibnamefont {Beser}}, \bibinfo
  {author} {\bibfnamefont {Y.}~\bibnamefont {Guan}}, \bibinfo {author}
  {\bibfnamefont {S.}~\bibnamefont {Kotochigova}}, \bibinfo {author}
  {\bibfnamefont {A.~M.}\ \bibnamefont {Lyyra}}, \bibinfo {author}
  {\bibfnamefont {S.}~\bibnamefont {Ashman}}, \bibinfo {author} {\bibfnamefont
  {C.~M.}\ \bibnamefont {Wolfe}}, \bibinfo {author} {\bibfnamefont
  {J.}~\bibnamefont {Huennekens}}, \bibinfo {author} {\bibfnamefont
  {F.}~\bibnamefont {Xie}}, \bibinfo {author} {\bibfnamefont {D.}~\bibnamefont
  {Li}}, \bibinfo {author} {\bibfnamefont {L.}~\bibnamefont {Li}}, \bibinfo
  {author} {\bibfnamefont {M.}~\bibnamefont {Tamanis}}, \bibinfo {author}
  {\bibfnamefont {R.}~\bibnamefont {Ferber}}, \bibinfo {author} {\bibfnamefont
  {A.}~\bibnamefont {Drozdova}}, \bibinfo {author} {\bibfnamefont
  {E.}~\bibnamefont {Pazyuk}}, \bibinfo {author} {\bibfnamefont {A.~V.}\
  \bibnamefont {Stolyarov}}, \bibinfo {author} {\bibfnamefont {J.~G.}\
  \bibnamefont {Danzl}}, \bibinfo {author} {\bibfnamefont {H.-C.}\ \bibnamefont
  {Naegerl}}, \bibinfo {author} {\bibfnamefont {N.}~\bibnamefont {Bouloufa}},
  \bibinfo {author} {\bibfnamefont {O.}~\bibnamefont {Dulieu}}, \bibinfo
  {author} {\bibfnamefont {C.}~\bibnamefont {Amiot}}, \bibinfo {author}
  {\bibfnamefont {H.}~\bibnamefont {Salami}},\ and\ \bibinfo {author}
  {\bibfnamefont {T.}~\bibnamefont {Bergeman}},\ }\href
  {https://doi.org/10.1103/PhysRevA.83.032514} {\bibfield  {journal} {\bibinfo
  {journal} {PHYSICAL REVIEW A}\ }\textbf {\bibinfo {volume} {83}},\ \bibinfo
  {pages} {032514} (\bibinfo {year} {2011})}\BibitemShut {NoStop}%
\bibitem [{\citenamefont {Znotins}\ \emph {et~al.}(2019)\citenamefont
  {Znotins}, \citenamefont {Kruzins}, \citenamefont {Tamanis}, \citenamefont
  {Ferber}, \citenamefont {Pazyuk}, \citenamefont {Stolyarov},\ and\
  \citenamefont {Zaitsevskii}}]{DP2CsCs2019}%
  \BibitemOpen
  \bibfield  {author} {\bibinfo {author} {\bibfnamefont {A.}~\bibnamefont
  {Znotins}}, \bibinfo {author} {\bibfnamefont {A.}~\bibnamefont {Kruzins}},
  \bibinfo {author} {\bibfnamefont {M.}~\bibnamefont {Tamanis}}, \bibinfo
  {author} {\bibfnamefont {R.}~\bibnamefont {Ferber}}, \bibinfo {author}
  {\bibfnamefont {E.~A.}\ \bibnamefont {Pazyuk}}, \bibinfo {author}
  {\bibfnamefont {A.}~\bibnamefont {Stolyarov}, \bibfnamefont {V}},\ and\
  \bibinfo {author} {\bibfnamefont {A.}~\bibnamefont {Zaitsevskii}},\ }\href
  {https://doi.org/10.1103/PhysRevA.100.042507} {\bibfield  {journal} {\bibinfo
   {journal} {PHYSICAL REVIEW A}\ }\textbf {\bibinfo {volume} {100}},\ \bibinfo
  {pages} {042507} (\bibinfo {year} {2019})}\BibitemShut {NoStop}%
\bibitem [{\citenamefont {Zaharova}\ \emph {et~al.}(2009)\citenamefont
  {Zaharova}, \citenamefont {Tamanis}, \citenamefont {Ferber}, \citenamefont
  {Drozdova}, \citenamefont {Pazyuk},\ and\ \citenamefont
  {Stolyarov}}]{DPNaCs2009}%
  \BibitemOpen
  \bibfield  {author} {\bibinfo {author} {\bibfnamefont {J.}~\bibnamefont
  {Zaharova}}, \bibinfo {author} {\bibfnamefont {M.}~\bibnamefont {Tamanis}},
  \bibinfo {author} {\bibfnamefont {R.}~\bibnamefont {Ferber}}, \bibinfo
  {author} {\bibfnamefont {A.~N.}\ \bibnamefont {Drozdova}}, \bibinfo {author}
  {\bibfnamefont {E.~A.}\ \bibnamefont {Pazyuk}},\ and\ \bibinfo {author}
  {\bibfnamefont {A.~V.}\ \bibnamefont {Stolyarov}},\ }\href
  {https://doi.org/10.1103/PhysRevA.79.012508} {\bibfield  {journal} {\bibinfo
  {journal} {PHYSICAL REVIEW A}\ }\textbf {\bibinfo {volume} {79}},\ \bibinfo
  {pages} {012508} (\bibinfo {year} {2009})}\BibitemShut {NoStop}%
\bibitem [{\citenamefont {Kruzins}\ \emph {et~al.}(2010)\citenamefont
  {Kruzins}, \citenamefont {Klincare}, \citenamefont {Nikolayeva},
  \citenamefont {Tamanis}, \citenamefont {Ferber}, \citenamefont {Pazyuk},\
  and\ \citenamefont {Stolyarov}}]{DP1KCs2010}%
  \BibitemOpen
  \bibfield  {author} {\bibinfo {author} {\bibfnamefont {A.}~\bibnamefont
  {Kruzins}}, \bibinfo {author} {\bibfnamefont {I.}~\bibnamefont {Klincare}},
  \bibinfo {author} {\bibfnamefont {O.}~\bibnamefont {Nikolayeva}}, \bibinfo
  {author} {\bibfnamefont {M.}~\bibnamefont {Tamanis}}, \bibinfo {author}
  {\bibfnamefont {R.}~\bibnamefont {Ferber}}, \bibinfo {author} {\bibfnamefont
  {E.~A.}\ \bibnamefont {Pazyuk}},\ and\ \bibinfo {author} {\bibfnamefont
  {A.~V.}\ \bibnamefont {Stolyarov}},\ }\href
  {https://doi.org/10.1103/PhysRevA.81.042509} {\bibfield  {journal} {\bibinfo
  {journal} {PHYSICAL REVIEW A}\ }\textbf {\bibinfo {volume} {81}},\ \bibinfo
  {pages} {042509} (\bibinfo {year} {2010})}\BibitemShut {NoStop}%
\bibitem [{\citenamefont {Tamanis}\ \emph {et~al.}(2010)\citenamefont
  {Tamanis}, \citenamefont {Klincare}, \citenamefont {Kruzins}, \citenamefont
  {Nikolayeva}, \citenamefont {Ferber}, \citenamefont {Pazyuk},\ and\
  \citenamefont {Stolyarov}}]{DP2KCs2010}%
  \BibitemOpen
  \bibfield  {author} {\bibinfo {author} {\bibfnamefont {M.}~\bibnamefont
  {Tamanis}}, \bibinfo {author} {\bibfnamefont {I.}~\bibnamefont {Klincare}},
  \bibinfo {author} {\bibfnamefont {A.}~\bibnamefont {Kruzins}}, \bibinfo
  {author} {\bibfnamefont {O.}~\bibnamefont {Nikolayeva}}, \bibinfo {author}
  {\bibfnamefont {R.}~\bibnamefont {Ferber}}, \bibinfo {author} {\bibfnamefont
  {E.~A.}\ \bibnamefont {Pazyuk}},\ and\ \bibinfo {author} {\bibfnamefont
  {A.~V.}\ \bibnamefont {Stolyarov}},\ }\href
  {https://doi.org/10.1103/PhysRevA.82.032506} {\bibfield  {journal} {\bibinfo
  {journal} {PHYSICAL REVIEW A}\ }\textbf {\bibinfo {volume} {82}},\ \bibinfo
  {pages} {032506} (\bibinfo {year} {2010})}\BibitemShut {NoStop}%
\bibitem [{\citenamefont {Kowalczyk}\ \emph {et~al.}(2015)\citenamefont
  {Kowalczyk}, \citenamefont {Jastrzebski}, \citenamefont {Szczepkowski},
  \citenamefont {Pazyuk},\ and\ \citenamefont {Stolyarov}}]{DPLiCs2015}%
  \BibitemOpen
  \bibfield  {author} {\bibinfo {author} {\bibfnamefont {P.}~\bibnamefont
  {Kowalczyk}}, \bibinfo {author} {\bibfnamefont {W.}~\bibnamefont
  {Jastrzebski}}, \bibinfo {author} {\bibfnamefont {J.}~\bibnamefont
  {Szczepkowski}}, \bibinfo {author} {\bibfnamefont {E.~A.}\ \bibnamefont
  {Pazyuk}},\ and\ \bibinfo {author} {\bibfnamefont {A.~V.}\ \bibnamefont
  {Stolyarov}},\ }\href {https://doi.org/10.1063/1.4922610} {\bibfield
  {journal} {\bibinfo  {journal} {JOURNAL OF CHEMICAL PHYSICS}\ }\textbf
  {\bibinfo {volume} {142}},\ \bibinfo {pages} {234308} (\bibinfo {year}
  {2015})}\BibitemShut {NoStop}%
\bibitem [{\citenamefont {Harker}\ \emph {et~al.}(2015)\citenamefont {Harker},
  \citenamefont {Crozet}, \citenamefont {Ross}, \citenamefont {Richter},
  \citenamefont {Jones}, \citenamefont {Faust}, \citenamefont {Huennekens},
  \citenamefont {Stolyarov}, \citenamefont {Salami},\ and\ \citenamefont
  {Bergeman}}]{DPNaK2015}%
  \BibitemOpen
  \bibfield  {author} {\bibinfo {author} {\bibfnamefont {H.}~\bibnamefont
  {Harker}}, \bibinfo {author} {\bibfnamefont {P.}~\bibnamefont {Crozet}},
  \bibinfo {author} {\bibfnamefont {A.~J.}\ \bibnamefont {Ross}}, \bibinfo
  {author} {\bibfnamefont {K.}~\bibnamefont {Richter}}, \bibinfo {author}
  {\bibfnamefont {J.}~\bibnamefont {Jones}}, \bibinfo {author} {\bibfnamefont
  {C.}~\bibnamefont {Faust}}, \bibinfo {author} {\bibfnamefont
  {J.}~\bibnamefont {Huennekens}}, \bibinfo {author} {\bibfnamefont {A.~V.}\
  \bibnamefont {Stolyarov}}, \bibinfo {author} {\bibfnamefont {H.}~\bibnamefont
  {Salami}},\ and\ \bibinfo {author} {\bibfnamefont {T.}~\bibnamefont
  {Bergeman}},\ }\href {https://doi.org/10.1103/PhysRevA.92.012506} {\bibfield
  {journal} {\bibinfo  {journal} {PHYSICAL REVIEW A}\ }\textbf {\bibinfo
  {volume} {92}},\ \bibinfo {pages} {012506} (\bibinfo {year}
  {2015})}\BibitemShut {NoStop}%
\bibitem [{\citenamefont {Alps}\ \emph {et~al.}(2016)\citenamefont {Alps},
  \citenamefont {Kruzins}, \citenamefont {Tamanis}, \citenamefont {Ferber},
  \citenamefont {Pazyuk},\ and\ \citenamefont {Stolyarov}}]{DPKRb2016}%
  \BibitemOpen
  \bibfield  {author} {\bibinfo {author} {\bibfnamefont {K.}~\bibnamefont
  {Alps}}, \bibinfo {author} {\bibfnamefont {A.}~\bibnamefont {Kruzins}},
  \bibinfo {author} {\bibfnamefont {M.}~\bibnamefont {Tamanis}}, \bibinfo
  {author} {\bibfnamefont {R.}~\bibnamefont {Ferber}}, \bibinfo {author}
  {\bibfnamefont {E.~A.}\ \bibnamefont {Pazyuk}},\ and\ \bibinfo {author}
  {\bibfnamefont {A.~V.}\ \bibnamefont {Stolyarov}},\ }\href
  {https://doi.org/10.1063/1.4945721} {\bibfield  {journal} {\bibinfo
  {journal} {JOURNAL OF CHEMICAL PHYSICS}\ }\textbf {\bibinfo {volume} {144}},\
  \bibinfo {pages} {144310} (\bibinfo {year} {2016})}\BibitemShut {NoStop}%
\bibitem [{\citenamefont {Press}\ \emph {et~al.}(2007)\citenamefont {Press},
  \citenamefont {Teukolsky}, \citenamefont {Vetterling},\ and\ \citenamefont
  {Flannery}}]{BookNumRec2007}%
  \BibitemOpen
  \bibfield  {author} {\bibinfo {author} {\bibfnamefont {W.~H.}\ \bibnamefont
  {Press}}, \bibinfo {author} {\bibfnamefont {S.~A.}\ \bibnamefont
  {Teukolsky}}, \bibinfo {author} {\bibfnamefont {W.~T.}\ \bibnamefont
  {Vetterling}},\ and\ \bibinfo {author} {\bibfnamefont {B.~P.}\ \bibnamefont
  {Flannery}},\ }\href {https://doi.org/10.1145/1874391.187410} {\emph
  {\bibinfo {title} {Numerical Recipes: The Art of Scientific Computing (3rd
  Edition)}}}\ (\bibinfo  {publisher} {Cambridge University Press},\ \bibinfo
  {year} {2007})\BibitemShut {NoStop}%
\bibitem [{\citenamefont {Marques}\ \emph {et~al.}(2008)\citenamefont
  {Marques}, \citenamefont {Prudente}, \citenamefont {Pereira}, \citenamefont
  {Almeida}, \citenamefont {Maniero},\ and\ \citenamefont
  {Fellows}}]{GANaLi2008}%
  \BibitemOpen
  \bibfield  {author} {\bibinfo {author} {\bibfnamefont {J.~M.~C.}\
  \bibnamefont {Marques}}, \bibinfo {author} {\bibfnamefont {F.~V.}\
  \bibnamefont {Prudente}}, \bibinfo {author} {\bibfnamefont {F.~B.}\
  \bibnamefont {Pereira}}, \bibinfo {author} {\bibfnamefont {M.~M.}\
  \bibnamefont {Almeida}}, \bibinfo {author} {\bibfnamefont {A.~M.}\
  \bibnamefont {Maniero}},\ and\ \bibinfo {author} {\bibfnamefont {C.~E.}\
  \bibnamefont {Fellows}},\ }\href
  {https://doi.org/10.1088/0953-4075/41/8/085103} {\bibfield  {journal}
  {\bibinfo  {journal} {JOURNAL OF PHYSICS B-ATOMIC MOLECULAR AND OPTICAL
  PHYSICS}\ }\textbf {\bibinfo {volume} {41}},\ \bibinfo {pages} {085103}
  (\bibinfo {year} {2008})}\BibitemShut {NoStop}%
\bibitem [{\citenamefont {Almeida}\ \emph {et~al.}(2011)\citenamefont
  {Almeida}, \citenamefont {Prudente}, \citenamefont {Fellows}, \citenamefont
  {Marques},\ and\ \citenamefont {Pereira}}]{GARbCs2011}%
  \BibitemOpen
  \bibfield  {author} {\bibinfo {author} {\bibfnamefont {M.~M.}\ \bibnamefont
  {Almeida}}, \bibinfo {author} {\bibfnamefont {F.~V.}\ \bibnamefont
  {Prudente}}, \bibinfo {author} {\bibfnamefont {C.~E.}\ \bibnamefont
  {Fellows}}, \bibinfo {author} {\bibfnamefont {J.~M.~C.}\ \bibnamefont
  {Marques}},\ and\ \bibinfo {author} {\bibfnamefont {F.~B.}\ \bibnamefont
  {Pereira}},\ }\href {https://doi.org/10.1088/0953-4075/44/22/225102}
  {\bibfield  {journal} {\bibinfo  {journal} {JOURNAL OF PHYSICS B-ATOMIC
  MOLECULAR AND OPTICAL PHYSICS}\ }\textbf {\bibinfo {volume} {44}},\ \bibinfo
  {pages} {225102} (\bibinfo {year} {2011})}\BibitemShut {NoStop}%
\bibitem [{\citenamefont {Stevenson}\ and\ \citenamefont
  {Pérez-Ríos}(2019)}]{GALiRb2019}%
  \BibitemOpen
  \bibfield  {author} {\bibinfo {author} {\bibfnamefont {I.~C.}\ \bibnamefont
  {Stevenson}}\ and\ \bibinfo {author} {\bibfnamefont {J.}~\bibnamefont
  {Pérez-Ríos}},\ }\href {https://doi.org/10.1088/1361-6455/ab0c4b}
  {\bibfield  {journal} {\bibinfo  {journal} {JOURNAL OF PHYSICS B-ATOMIC
  MOLECULAR AND OPTICAL PHYSICS}\ }\textbf {\bibinfo {volume} {52}},\ \bibinfo
  {pages} {105002} (\bibinfo {year} {2019})}\BibitemShut {NoStop}%
\bibitem [{\citenamefont {Londo\~no}\ and\ \citenamefont
  {Arce}(2023)}]{GALD2023}%
  \BibitemOpen
  \bibfield  {author} {\bibinfo {author} {\bibfnamefont {M.}~\bibnamefont
  {Londo\~no}}\ and\ \bibinfo {author} {\bibfnamefont {J.~C.}\ \bibnamefont
  {Arce}},\ }\href {https://doi.org/10.1103/PhysRevA.108.013301} {\bibfield
  {journal} {\bibinfo  {journal} {Phys. Rev. A}\ }\textbf {\bibinfo {volume}
  {108}},\ \bibinfo {pages} {013301} (\bibinfo {year} {2023})}\BibitemShut
  {NoStop}%
\bibitem [{\citenamefont {Katoch}\ \emph {et~al.}(2021)\citenamefont {Katoch},
  \citenamefont {Chauhan},\ and\ \citenamefont {Kumar}}]{RevGA2021}%
  \BibitemOpen
  \bibfield  {author} {\bibinfo {author} {\bibfnamefont {S.}~\bibnamefont
  {Katoch}}, \bibinfo {author} {\bibfnamefont {S.~S.}\ \bibnamefont
  {Chauhan}},\ and\ \bibinfo {author} {\bibfnamefont {V.}~\bibnamefont
  {Kumar}},\ }\href {https://doi.org/10.1007/s11042-020-10139-6} {\bibfield
  {journal} {\bibinfo  {journal} {MULTIMEDIA TOOLS AND APPLICATIONS}\ }\textbf
  {\bibinfo {volume} {80}},\ \bibinfo {pages} {8091} (\bibinfo {year}
  {2021})}\BibitemShut {NoStop}%
\bibitem [{\citenamefont {Hua}(1990)}]{WeiPOT1990}%
  \BibitemOpen
  \bibfield  {author} {\bibinfo {author} {\bibfnamefont {W.}~\bibnamefont
  {Hua}},\ }\href {https://doi.org/10.1103/PhysRevA.42.2524} {\bibfield
  {journal} {\bibinfo  {journal} {Phys. Rev. A}\ }\textbf {\bibinfo {volume}
  {42}},\ \bibinfo {pages} {2524} (\bibinfo {year} {1990})}\BibitemShut
  {NoStop}%
\bibitem [{NIS()}]{NISTDatabase}%
  \BibitemOpen
  \href {https://physics.nist.gov/PhysRefData/ASD/levels_form.html} {\bibinfo
  {title} {Nist: Atomic spectra database}},\ \bibinfo {note} {accessed on May
  10th, 2024 [URL will be inserted by publisher]}\BibitemShut {NoStop}%
\bibitem [{\citenamefont {Yin}\ \emph {et~al.}(2024)\citenamefont {Yin},
  \citenamefont {Li}, \citenamefont {Bai}, \citenamefont {Gong}, \citenamefont
  {Ji}, \citenamefont {Zhao}, \citenamefont {Han}, \citenamefont {Yu},\ and\
  \citenamefont {Wang}}]{GAYin2024}%
  \BibitemOpen
  \bibfield  {author} {\bibinfo {author} {\bibfnamefont {Y.}~\bibnamefont
  {Yin}}, \bibinfo {author} {\bibfnamefont {Z.}~\bibnamefont {Li}}, \bibinfo
  {author} {\bibfnamefont {X.}~\bibnamefont {Bai}}, \bibinfo {author}
  {\bibfnamefont {T.}~\bibnamefont {Gong}}, \bibinfo {author} {\bibfnamefont
  {Z.}~\bibnamefont {Ji}}, \bibinfo {author} {\bibfnamefont {Y.}~\bibnamefont
  {Zhao}}, \bibinfo {author} {\bibfnamefont {Y.}~\bibnamefont {Han}}, \bibinfo
  {author} {\bibfnamefont {J.}~\bibnamefont {Yu}},\ and\ \bibinfo {author}
  {\bibfnamefont {G.}~\bibnamefont {Wang}},\ }\href
  {https://doi.org/10.1088/1402-4896/ad2b3e} {\bibfield  {journal} {\bibinfo
  {journal} {Physica Scripta}\ }\textbf {\bibinfo {volume} {99}},\ \bibinfo
  {pages} {045003} (\bibinfo {year} {2024})}\BibitemShut {NoStop}%
\bibitem [{Note1()}]{Note1}%
  \BibitemOpen
  \bibinfo {note} {Since the energy range of the observed rovibrational levels
  is limited, the crossover points should also be limited in certain
  regions}\BibitemShut {NoStop}%
\bibitem [{\citenamefont {Kraft}(1988)}]{SLSQP1988}%
  \BibitemOpen
  \bibfield  {author} {\bibinfo {author} {\bibfnamefont {D.}~\bibnamefont
  {Kraft}},\ }\href@noop {} {\emph {\bibinfo {title} {A software package for
  sequential quadratic programming}}},\ \bibinfo {type} {Tech. Rep.}\ (\bibinfo
   {institution} {DFVLR-FB 88-28, DLR German Aerospace Center – Institute for
  Flight Mechanics, Koln, Germany.},\ \bibinfo {year} {1988})\BibitemShut
  {NoStop}%
\bibitem [{\citenamefont {Le~Roy}\ and\ \citenamefont
  {Pashov}(2017)}]{betaFIT2017}%
  \BibitemOpen
  \bibfield  {author} {\bibinfo {author} {\bibfnamefont {R.~J.}\ \bibnamefont
  {Le~Roy}}\ and\ \bibinfo {author} {\bibfnamefont {A.}~\bibnamefont
  {Pashov}},\ }\href {https://doi.org/10.1016/j.jqsrt.2016.03.036} {\bibfield
  {journal} {\bibinfo  {journal} {JOURNAL OF QUANTITATIVE SPECTROSCOPY \&
  RADIATIVE TRANSFER}\ }\textbf {\bibinfo {volume} {186}},\ \bibinfo {pages}
  {210} (\bibinfo {year} {2017})}\BibitemShut {NoStop}%
\bibitem [{\citenamefont {MARSTON}\ and\ \citenamefont
  {BALINTKURTI}(1989)}]{FGH1989}%
  \BibitemOpen
  \bibfield  {author} {\bibinfo {author} {\bibfnamefont {C.}~\bibnamefont
  {MARSTON}}\ and\ \bibinfo {author} {\bibfnamefont {G.}~\bibnamefont
  {BALINTKURTI}},\ }\href {https://doi.org/10.1063/1.456888} {\bibfield
  {journal} {\bibinfo  {journal} {JOURNAL OF CHEMICAL PHYSICS}\ }\textbf
  {\bibinfo {volume} {91}},\ \bibinfo {pages} {3571} (\bibinfo {year}
  {1989})}\BibitemShut {NoStop}%
\bibitem [{opt()}]{optSciPy}%
  \BibitemOpen
  \href
  {https://docs.scipy.org/doc/scipy/reference/optimize.minimize-slsqp.html}
  {\bibinfo {title} {Scipy v1.13.0 manual}},\ \bibinfo {note} {accessed on May
  10th, 2024 [URL will be inserted by publisher]}\BibitemShut {NoStop}%
\bibitem [{Note2()}]{Note2}%
  \BibitemOpen
  \bibinfo {note} {See Supplemental Material at [URL will be inserted by
  publisher].}\BibitemShut {Stop}%
\bibitem [{\citenamefont {Cairncross}\ \emph {et~al.}(2021)\citenamefont
  {Cairncross}, \citenamefont {Zhang}, \citenamefont {Picard}, \citenamefont
  {Yu}, \citenamefont {Wang},\ and\ \citenamefont {Ni}}]{MANaCsNi2021}%
  \BibitemOpen
  \bibfield  {author} {\bibinfo {author} {\bibfnamefont {W.~B.}\ \bibnamefont
  {Cairncross}}, \bibinfo {author} {\bibfnamefont {J.~T.}\ \bibnamefont
  {Zhang}}, \bibinfo {author} {\bibfnamefont {L.~R.~B.}\ \bibnamefont
  {Picard}}, \bibinfo {author} {\bibfnamefont {Y.}~\bibnamefont {Yu}}, \bibinfo
  {author} {\bibfnamefont {K.}~\bibnamefont {Wang}},\ and\ \bibinfo {author}
  {\bibfnamefont {K.-K.}\ \bibnamefont {Ni}},\ }\href
  {https://doi.org/10.1103/PhysRevLett.126.123402} {\bibfield  {journal}
  {\bibinfo  {journal} {Phys. Rev. Lett.}\ }\textbf {\bibinfo {volume} {126}},\
  \bibinfo {pages} {123402} (\bibinfo {year} {2021})}\BibitemShut {NoStop}%
\bibitem [{\citenamefont {Christakis}\ \emph {et~al.}(2023)\citenamefont
  {Christakis}, \citenamefont {Rosenberg}, \citenamefont {Raj}, \citenamefont
  {Chi}, \citenamefont {Morningstar}, \citenamefont {Huse}, \citenamefont
  {Yan},\ and\ \citenamefont {Bakr}}]{MANa23Rb872023}%
  \BibitemOpen
  \bibfield  {author} {\bibinfo {author} {\bibfnamefont {L.}~\bibnamefont
  {Christakis}}, \bibinfo {author} {\bibfnamefont {J.~S.}\ \bibnamefont
  {Rosenberg}}, \bibinfo {author} {\bibfnamefont {R.}~\bibnamefont {Raj}},
  \bibinfo {author} {\bibfnamefont {S.}~\bibnamefont {Chi}}, \bibinfo {author}
  {\bibfnamefont {A.}~\bibnamefont {Morningstar}}, \bibinfo {author}
  {\bibfnamefont {D.~A.}\ \bibnamefont {Huse}}, \bibinfo {author}
  {\bibfnamefont {Z.~Z.}\ \bibnamefont {Yan}},\ and\ \bibinfo {author}
  {\bibfnamefont {W.~S.}\ \bibnamefont {Bakr}},\ }\href
  {https://doi.org/10.1038/s41586-022-05558-4} {\bibfield  {journal} {\bibinfo
  {journal} {Nature}\ }\textbf {\bibinfo {volume} {614}},\ \bibinfo {pages}
  {64} (\bibinfo {year} {2023})}\BibitemShut {NoStop}%
\end{thebibliography}%

\end{document}